\documentclass[times, twocolumn]{aastex63}
\usepackage{amsfonts,amsmath,amssymb,amsthm,multirow}
\usepackage{booktabs}
\usepackage{textcomp}
\usepackage{natbib}
\usepackage[flushleft]{threeparttable}
\usepackage{lipsum}
\usepackage{makecell}

\providecommand{\e}[1]{\ensuremath{\times 10^{#1}}} 

\newcommand{\teff}{$T_{\rm eff}$}
\setcellgapes{0pt}

\date{\today}

\begin{document}

\shortauthors{Zalesky et al.}

\title{A Uniform Retrieval Analysis of Ultra-cool Dwarfs. IV. A Statistical Census from 50 Late-T Dwarfs}
\correspondingauthor{Joe Zalesky}
\email{jazalesk@asu.edu}
\author[0000-0001-9828-1848]{Joseph A. Zalesky}
\affil{School of Earth \& Space Exploration, Arizona State University, Tempe AZ 85287, USA}
\author[0000-0002-9967-3752]{Kezman Saboi}
\affil{School of Earth \& Space Exploration, Arizona State University, Tempe AZ 85287, USA}
\author[0000-0002-2338-476X]{Michael R. Line}
\affil{School of Earth \& Space Exploration, Arizona State University, Tempe AZ 85287, USA}
\author[0000-0002-3726-4881]{Zhoujian Zhang}
\affil{Institute for Astronomy, University of Hawaii, 2680 Woodlawn Drive, Honolulu, HI 96822, USA}
\author[0000-0002-6294-5937]{Adam C. Schneider}
\affiliation{US Naval Observatory, Flagstaff Station, P.O. Box 1149, Flagstaff, AZ 86002, USA}
\affiliation{Department of Physics and Astronomy, George Mason University, MS3F3, 4400 University Drive, Fairfax, VA 22030, USA}
\author[0000-0003-2232-7664]{Michael C. Liu}
\affil{Institute for Astronomy, University of Hawaii, 2680 Woodlawn Drive, Honolulu, HI 96822, USA}
\author[0000-0003-0562-1511]{William M. J. Best}
\affil{University of Texas at Austin, Department of Astronomy, 2515 Speedway C1400, Austin, TX 78712, USA}
\author[0000-0002-5251-2943]{Mark S. Marley}
\affil{Lunar \& Planetary Laboratory, University of Arizona, Tucson, AZ 85721}

\begin{abstract}
The spectra of brown dwarfs are key to exploring the chemistry and physics that take place in their atmospheres. Late-T dwarf spectra are particularly diagnostic due to their relatively cloud-free atmospheres and deep molecular bands. With the use of powerful atmospheric retrieval tools applied to the spectra of these objects, direct constraints on molecular/atomic abundances, gravity, and vertical thermal profiles can be obtained enabling a broad exploration of the chemical/physical mechanisms operating in their atmospheres. We present a uniform retrieval analysis on low-resolution IRTF SpeX near-IR spectra of a sample of 50 T dwarfs, including new observations as part of a recent volume-limited survey. This analysis more than quadruples the sample of T dwarfs with retrieved temperature profiles and abundances (H$_2$O, CH$_4$, NH$_3$, K and subsequent C/O and metallicities). We are generally able to constrain effective temperatures to within 50K, volume mixing ratios for major species to within 0.25dex, atmospheric metallicities [M/H] to within 0.2, and C/O ratios to within 0.2. We compare our retrieved constraints on the thermal structure, chemistry, and gravities of these objects with predictions from self-consistent radiative-convective equilibrium models and find, in general though with substantial scatter, consistency with solar composition chemistry and thermal profiles of the neighboring stellar FGK population. Objects with notable discrepancies between the two modeling techniques and potential mechanisms for their differences, be they related to modeling approach or physically motivated, are discussed more thoroughly in the text.
\end{abstract}


\section{Introduction} \label{sec:intro}

Brown dwarfs are substellar objects whose masses are intermediate between the latest M-type stars and the most massive planets \citep{1963PThPh..30..460H, 1977ApJ...214..488S, 1988Natur.336..656B, 1995Natur.377..129R, 1995Sci...270.1478O, 2006ApJ...647..552S}.  Similar to stars, brown dwarfs form from interstellar molecular gas cloud core collapse \citep{2000ApJ...531L..91U, 2002MNRAS.332L..65B, 2005Natur.438..332K, Whitworth_2006, 2007prpl.conf..623C, 2007prpl.conf..459W, 2012EAS....57...91H}, but do not achieve masses high enough to sustain core H-fusion over their lifetime \citep{2001RvMP...73..719B}. As the effective temperatures of brown dwarfs are much cooler than those of stars ($<2500$K), molecules and condensates form in their photospheres and dominate the spectral energy distribution. It is through analyses of these spectra that we are able to infer the nature of brown dwarf atmospheres and how they evolve overtime. Hence we are able to use them as unirradiated laboratories, which provides a valuable baseline for understanding extrasolar planet atmospheres.

Fundamental properties of brown dwarfs include their mass, radius, gravity, effective temperatures (\teff), and elemental abundances (see review in \citealt{2005ARA&A..43..195K, 2015ARA&A..53..279M}). Robust masses can be derived through dynamical means \citep{2017ApJS..231...15D} and radius is inferred via distance and bolometric luminosity. Determining gravity, effective temperatures, and abundances for brown dwarfs can be more challenging than it is for stars due to the lack of clear atomic lines for which classic single or multi-line spectral analyses can be performed.  As such, atmospheric models which properly incorporate molecules and condensate species play a more important role in these determinations.

A common modeling approach for determining these properties is through comparisons of observations to grids of model spectra computed with self-consistent one-dimensional radiative-convective equilibrium models (e.g., \citealt{1996ApJ...465L.123A, 1996Sci...272.1919M, 2001RvMP...73..719B}) or using more recent grid-fitting methods \citep{2021ApJ...916...53Z, 2021arXiv210505256Z}. This approach typically relies upon the use of a priori chemical/physical assumptions such as thermochemical equilibrium, molecular/atomic abundances, assumed atmospheric chemistry paradigms, and 1D radiative-convective equilibrium \citep{2001RvMP...73..719B, 2015ARA&A..53..279M}. These assumptions are made to reduce the dimensionality of the inference problem to just a handful of parameters such as \teff, log($g$), and a composition parameter like metallicity $[M/H]$ \citep[in some cases, alpha-element enhancement e.g.][]{2013A&A...553A...6H}.

A few key issues arise with this method. First, the low model dimensionality restricts any inference solely to the dimensions specified for the pre-computed grid. Second, the choice of inference tool is often not rigorous and typically does not account for grid-interpolation uncertainties (e.g., often a simple chi-square type minimizer is combined with a multi-linear-type interpolater) and can result in artificially precise constraints. Thirdly, often times the overly restrictive assumptions lead to poor model spectra fits to the data (e.g., \cite{2012A&A...540A..85P}), leading one to question the validity of the self-consistent modeling assumptions.  \cite{2021ApJ...916...53Z, 2021arXiv210505256Z} sought to remedy the second and third issues through the use of a modernized self-consistent grid (\cite{2021arXiv210707434M}) combined with the {\tt Starfish} tool (\cite{2015ascl.soft05007C}), which attempts to account for the finite model grid spacing, interpolation uncertainties, and data-model misfits. However, such an approach is still restricted to a priori physical assumptions within the grid itself.

A more flexible alternative known as atmospheric retrieval, has shown success in providing robust model-data fits to low resolution spectral observations of T and L dwarfs \citep{2014ApJ...793...33L,2015ApJ...807..183L, burningham2017retrieval, 2020ApJ...905...46G}. Originally developed to determine temperatures and abundances from spectral soundings of the Solar System planets \citep{2007Icar..189..457F, 2008JQSRT.109.1136I, 2011Icar..214..606G}, this technique relies on the use of a forward radiative transfer model that relaxes many of the self-consistent grid model assumptions at the expense of requiring many free parameters. The fundamental philosophy of the retrieval approach, in contrast to the grid approach, is that much of the physical/chemical mechanisms operating in low temperature atmospheres are not understood well enough to build accurate, fully self-consistent models. The aim of atmospheric retrieval is to directly determine, from the spectra the vertical temperature profiles and atmospheric composition.  This approach has recently become prolific in the extrasolar planet atmosphere studies (e.g., see review by \cite{2019ASPC..519..129M}).

\cite{2015ApJ...807..183L} applied the atmospheric retrieval approach to low resolution SpeX data of the benchmark late-T dwarfs, Gl 570D and HD 3651B. Late-T dwarfs were specifically chosen to mitigate the impact of uncertain cloud properties and the presentation of strong molecular absorption features from water and methane.  Using a radiative transfer forward model with $\sim$30 free parameters combined with Markov Chain Monte Carlo (MCMC) inference \citep{2013PASP..125..306F}, they were able to obtain bound constraints on the molecular mixing ratios for [H$_2$O, CH$_4$, NH$_3$ and Na+K], the vertical thermal profiles (temperature vs. pressure), gravity, and photometric radii (given the parallactic distances).  The key findings were (1) that ammonia abundance could be constrained from low-resolution near-infrared spectra alone, a surprise given the lack of obvious visual spectral features (typical of longer wavelengths \citealt{2006ApJ...647..552S} or higher resolutions \citep{2015MNRAS.450..454C}); (2) the retrieved molecular (and alkali) abundances were consistent with self-consistent chemical predictions (albeit constant-with-altitude mixing ratios were assumed in the retrieval); (3) derived metallicities and carbon-to-oxygen ratios were consistent with their host star abundances; and (4) the vertical thermal temperature profiles agreed with radiative-convective equilibrium. Taken together, these findings lend support that the retrieval paradigm can be used as a complimentary tool to grid-modeling for inferring fundamental brown dwarf atmospheric properties.  

Having validated the retrieval methodology against late-T benchmark systems,  \cite{2017ApJ...848...83L} performed a systematic retrieval analysis on the spectra of 11 late-T dwarfs (T7-T8, spanning 600 - 800K) available in the SpeX prism library \citep{2014ASInC..11....7B}. This uniform analysis found that (1) the large number of free parameters required in retrievals, compared to self-consistent grid models (27 vs. 4 parameters) is justified owing to their much better fits; (2) the T7/T8 atmospheres are cloud free (upper limits on the cloud optical depth of unity were obtained); (3) the temperature profiles for all objects were again consistent with radiative-convective equilibrium; (4) the retrieved gravities, radii, and inferred effective temperatures agreed with evolution model predictions;  (5)  abundances for ammonia, methane, and water were found to be constant with effective temperature but a strong decreasing trend in the alkali abundances was observed to occur with decreasing effective temperature; and (6) the late-T dwarf ensemble had somewhat lower metallicities and higher carbon-to-oxygen ratios than the local FGK stellar population. These findings provided a first look at the carbon-to-oxygen ratio and metallicities of a sample brown dwarfs as well as the first direct determination of the possible influence of alkali rainout on their abundances. 

Building upon \cite{2015ApJ...807..183L, 2017ApJ...848...83L}, \cite{2019ApJ...877...24Z} extended the uniform retrieval analysis into the cooler Y-dwarfs. This sampled comprised eleven Y dwarfs and three T dwarfs as observed with the {\it Hubble Space Telescope}'s Wide Field Camera 3 (WFC3)  \citep{2015ApJ...804...92S}. The conclusions were similar to the late-T results \cite{2017ApJ...848...83L}, finding that (1) the retrieved temperature profiles for most objects were consistent with radiative-convective equilibrium predictions; (2) water and methane abundances were consistent with equilibrium chemistry; (3) the ammonia abundances showed an upward rise with decreasing temperature, with scatter consistent with both equilibrium and quenched abundances; (4) constraints and/or upper limits on the alkali abundances consistent with rain-out predictions; (5) findings of very high gravities, pushing the limits of the evolution models; and (6) elemental abundance ratios broadly inline with those from the local FGK star population. They also compared the results of a self-consistent grid fit to the retrieval results finding that (1) the retrievals fit better and the large number of free parameters were justified, and (2) constraints on common parameters (\teff, log($g$), metallicity, and radius) were inconsistent at 1$\sigma$.

Additional works by \cite{burningham2017retrieval} and \cite{2020ApJ...905...46G} focused on determining the fundamental properties of L-dwarfs using similar retrieval methods. L-dwarfs are more complicated to characterize due to the presence of condensate clouds and additional higher temperature species (hydrides, oxides), and the reduced vertical grasp on the thermal structures.  Overall these cloudy investigations showed that the retrieved temperature structures could be degenerate with the presence of clouds, but that plausible abundances of the hydrides/oxides could be retrieved, opening up the possibility of abundance determinations at higher temperatures.

In this work, we extend the late-T analysis in \cite{2017ApJ...848...83L} to a broader sample of 50 T7-T9-dwarfs which were contained in a volume limited survey (\cite{2020AJ....159..257B}), including new IRTF SpeX spectroscopy \citep{2021ApJ...916...53Z, 2021arXiv210505256Z}. Our work here represents the largest uniform retrieval analysis on a nearly complete sample of late-T dwarfs, with a factor of 5 larger sample over the analysis in \cite{2017ApJ...848...83L}. As in the past papers of this series, we focus on late-T dwarfs as these have been shown to be largely free of influence from clouds and they present deep methane and water absorption features, enabling simultaneous carbon and oxygen constraints. 

We follow closely the methodologies and analysis as in \cite{2017ApJ...848...83L} and \cite{2019ApJ...877...24Z}. In Section \ref{sec:Methods}, we briefly discuss the source of our data and review our retrieval methodology. Section \ref{sec: results} presents our retrieved constraints for our thermal structures (\ref{sec:Thermals}), composition (\ref{sec: Atmospheric composition}), and evolutionary parameters (\ref{sec: Basics physical properties}). Finally we briefly summarize our findings in Section \ref{sec: summary}.

\section{Data} \label{sec:Data}
\begin{figure}[b]
    \centering
    \includegraphics[width=\columnwidth]{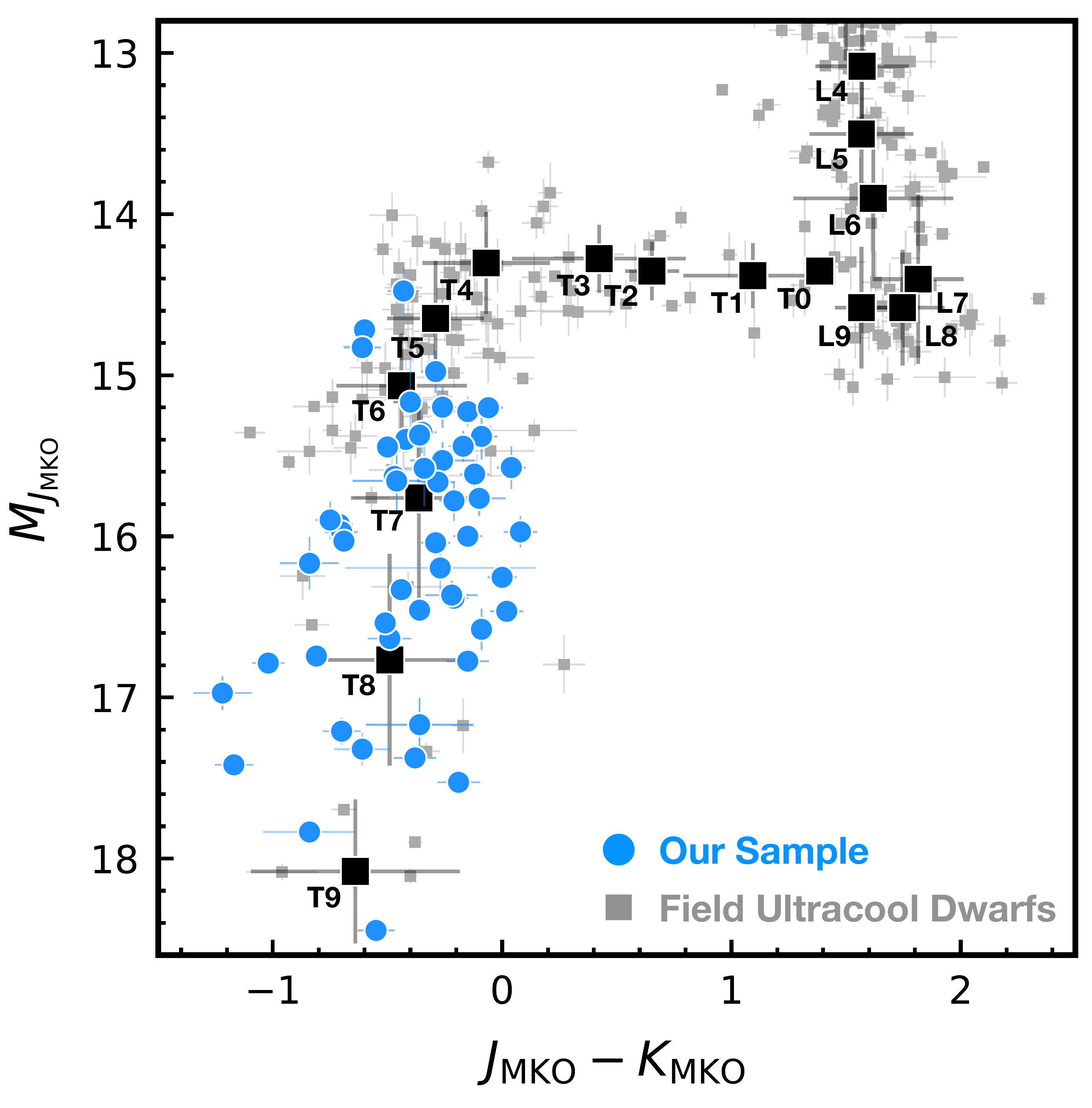}
    \caption{A color-magnitude diagram of a population of field brown dwarfs \citep{2018ApJS..234....1B, 2020AJ....159..257B}, and our sample of brown dwarfs from this study (T5-T9). This collective ensemble shows L and T dwarfs, in descending order from the top of the diagram to the bottom.
    }
    \label{fig:color}
\end{figure}
\begin{table*}[ht]   
    \centering
    \scriptsize 
    \caption{Basic properties of the 50 late-T dwarfs used in our retrieval analysis.}
    \begin{threeparttable}
    \begin{tabular}{c c c c c c c}
        \toprule
        \small Name of Object & \small Spec. Type & \small $J_{\rm MKO}$ [mag] & \small $H_{\rm MKO}$ [mag] & \small $K_{\rm MKO}$ [mag] &  \small  Distance [pc] & \small J-Band Peak SNR\\
        \midrule
        HD3651B & T7.5 & 16.16$^{+0.03}_{-0.03}$ & 16.68$^{+0.04}_{-0.04}$ & 16.87$^{+0.05}_{-0.05}$ & 11.14$^{+0.01}_{-0.01}$ \textbf{(4)} & 73.7\\
        WISE J004024.88+090054.8 & T7 & 16.13$^{+0.01}_{-0.01}$ & 16.56$^{+0.02}_{-0.02}$ & 16.55$^{+0.05}_{-0.05}$ & 14.01$^{+0.53}_{-0.53}$ \textbf{(1)} & 73.4\\
        WISE J004945.61+215120.0PRZ0.5 & T8 & 17.63$^{+0.13}_{-0.13}$ & 18.09$^{+0.14}_{-0.14}$ &  18.06$^{+0.14}_{-0.14}$ & 24.81$^{+1.48}_{-1.48}$ \textbf{(3)} & 76.5\\
        2MASS J00501994-3322402 & T7 & 15.65$^{+0.1}_{-0.1}$ &16.04 $^{+0.1}_{-0.1}$ & 15.91$^{+0.1}_{-0.1}$ & 10.57$^{+0.27}_{-0.27}$ \textbf{(2)} & 60.4\\
        WISEPA J012333.21+414203.9 & T7 & 17.00$^{+0.02}_{-0.02}$ & 17.29$^{+0.06}_{-0.06}$ & 17.29$^{+0.06}_{-0.06}$ & 25.38$^{+1.55}_{-1.55}$ \textbf{(1)} & 67.7\\
        WISEPC J022322.39-293258.1&  T7.5& 17.1$^{+0.05}_{-0.05}$ & 17.3$^{+0.11}_{-0.11}$ & 17.59$^{+0.08}_{-0.08}$ & 12.39$^{+0.4}_{-0.4}$ \textbf{(3)} & 58.3\\
        WISE J024124.73-365328.0 &  T7 & 16.59$^{+0.04}_{-0.04}$ & 17.04$^{+0.07}_{-0.07}$ & N/A & 19.08$^{+0.98}_{-0.98}$ \textbf{(3)} & 66.1\\
        PSO J043.5395+02.3995& T8 & 15.92$^{+0.01}_{-0.01}$ & 16.29$^{+0.02}_{-0.02}$ & 16.73$^{+0.05}_{-0.05}$ & 6.84$^{+0.07}_{-0.07}$ \textbf{(3)} & 125.0\\
        WISE J032547.72+083118.2 &  T7 & 16.29$^{+0.07}_{-0.07}$ & 16.19$^{+0.08}_{-0.08}$ & 16.39$^{+0.09}_{-0.09}$ & 12.74$^{+0.49}_{-0.49}$ \textbf{(3)} & 66.1\\
        2MASSI J0415195-093506 & T8 & 15.32$^{+0.03}_{-0.03}$ & 15.7$^{+0.03}_{-0.03}$ & 15.83$^{+0.03}_{-0.03}$ & 5.71$^{+0.06}_{-0.06}$ \textbf{(2)} & 223.0\\
        WISEPA J045853.89+643452.9& T8.5 & 17.13$^{+0.07}_{-0.07}$ & 17.45$^{+0.11}_{-0.11}$ & 17.74$^{+0.1}_{-0.1}$ & 9.16$^{+0.3}_{-0.3}$ \textbf{(3)} & 35.5\\
        WISE J052126.29+102528.4&  T7.5& 14.86$^{+0.02}_{-0.02}$ & 15.25$^{+0.06}_{-0.06}$ & 14.98$^{+0.06}_{-0.06}$ & 7.07$^{+0.25}_{-0.25}$ \textbf{(1)} & 93.8\\
        UGPS J052127.27+364048.6 & T8.5 & 16.94$^{+0.02}_{-0.02}$ & 17.32$^{+0.09}_{-0.09}$& 17.28$^{+0.04}_{-0.04}$& \phn8.18$^{+0.11}_{-0.11}$ \textbf{(3)} & 60.5\\
        WISE J061437.73+095135.0& T7 & 16.43$^{+0.02}_{-0.02}$ & 16.64$^{+0.06}_{-0.06}$ & 16.49$^{+0.06}_{-0.06}$ & 17.61$^{+0.62}_{-0.62}$ \textbf{(1)} & 106.1\\
        2MASSI J0727182+171001& T7 & 15.19$^{+0.03}_{-0.03}$ & 15.67$^{+0.03}_{-0.03}$ & 16.69$^{+0.03}_{-0.03}$ & 8.89$^{+0.07}_{-0.07}$ \textbf{(2)} & 47.8\\
        2MASS J07290002-3954043 & T8pec & 15.66$^{+0.08}_{-0.08}$ & 16.05$^{+0.1}_{-0.1}$ & 16.5$^{+0.1}_{-0.1}$ & 7.92$^{+0.52}_{-0.52}$ \textbf{(8)} & 58.6\\
        2MASS J09393548-2448279 & T8 & 15.61$^{+0.09}_{-0.09}$ & 15.96$^{+0.09}_{-0.09}$ & 16.83$^{+0.09}_{-0.09}$ & 5.34$^{+0.13}_{-0.13}$ \textbf{(7)} & 64.0\\
        ULAS J102940.52+093514.6& T8 & 17.28$^{+0.01}_{-0.01}$ & 17.63$^{+0.01}_{-0.01}$ & 17.64$^{+0.02}_{-0.02}$ & 14.6$^{+0.36}_{-0.36}$ \textbf{(1)} & 54.1\\
        WISE J103907.73-160002.9 & T7.5 & 16.95$^{+0.02}_{-0.02}$ & 17.19$^{+0.04}_{-0.04}$ & 17.1$^{+0.07}_{-0.07}$ & 22.12$^{+0.93}_{-0.93}$ \textbf{(1)} & 35.6\\
        WISE J105257.95-194250.2 & T7.5 & 16.84$^{+0.02}_{-0.02}$  & 16.99$^{+0.06}_{-0.06}$ & 17.07$^{+0.06}_{-0.06}$ & 14.73$^{+0.48}_{-0.48}$ \textbf{(1)} & 71.4\\
        2MASS J11145133-2618235 & T7.5 & 15.52$^{+0.05}_{0.05-}$ & 15.82$^{+0.05}_{-0.05}$ & 16.54$^{+0.05}_{-0.05}$ & 5.58$^{+0.04}_{-0.04}$ \textbf{(2)} & 26.0\\
        WISE J112438.12-042149.7 & T7 & 16.72$^{+0.13}_{-0.13}$ & 16.37 & 16.32  & 17.39$^{+0.10}_{-0.10}$ \textbf{(1)} & 85.4\\
        2MASSI J1217110-031113 & T7.5 & 15.56$^{+0.03}_{-0.03}$ & 15.98$^{+0.03}_{-0.03}$ & 15.92$^{+0.03}_{-0.03}$ & 10.91$^{+0.26}_{-0.26}$ \textbf{(5),(3)} & 31.0\\
        WISE J125448.52-072828.4& T8 & 17.3$^{+0.01}_{-0.01}$ & 17.63$^{+0.03}_{-0.03}$ & 17.39$^{+0.07}_{-0.07}$ & 24.21$^{+1.58}_{-1.58}$ \textbf{(1)} & 25.0\\
        WISE J125715.90+400854.2&  T7& 16.88$^{+0.02}_{-0.02}$ & 17.12$^{+0.06}_{-0.06}$ & 17.16$^{+0.07}_{-0.07}$ & 17.51$^{+0.55}_{-0.55}$ \textbf{(1)} & 32.9\\
        Ross 458C& T8 & 16.69$^{+0.02}_{-0.02}$ & 17.01$^{+0.04}_{-0.04}$ & 16.9$^{+0.06}_{-0.06}$ & 11.51$^{+0.02}_{-0.02}$ \textbf{(4)} & 41.5\\
        WISEPA J132233.66-234017.1& T8 & 16.75$^{+0.11}_{-0.11}$ & 16.65$^{+0.14}_{-0.14}$ & 17.02 $^{+0.4}_{-0.4}$ & 12.9$^{+0.7}_{-0.7}$ \textbf{(3)} & 31.5\\
        ULAS J141623.94+134836.3 & (sd)T7.5 & 17.26$^{+0.02}_{-0.02}$ & 17.58$^{+0.03}_{-0.03}$ & 18.43$^{+0.08}_{-0.08}$ & 9.3$^{+0.03}_{-0.03}$ \textbf{(4)} & 52.0\\
        WISEPC J145715.03+581510.2&  T7& 16.82$^{+0.02}_{-0.02}$ & 17.16$^{+0.06}_{-0.06}$ & 17.22$^{+0.07}_{-0.07}$ & 21.41$^{+2.61}_{-2.61}$ \textbf{(1)} & 51.8\\
        Gliese 570D & T7.5 & 14.82$^{+0.05}_{-0.05}$ & 15.28$^{+0.05}_{-0.05}$ & 15.52$^{+0.05}_{-0.05}$ & 5.88 \textbf{(4)} & 31.7\\
        PSO J224.3820+47.4057 & T7 &17.1$^{+0.02}_{-0.02}$ & 17.43 $^{+0.06}_{-0.06}$ & 17.06$^{+0.06}_{-0.06}$ & 20.2$^{+1.22}_{-1.22}$ \textbf{(1)} & 34.4\\
        SDSS J150411.63+102718.4 & T7 & 16.51$^{+0.01}_{-0.01}$ & 16.99$^{+0.05}_{-0.05}$ & 17.12$^{+0.08}_{-0.08}$ & 21.69$^{+0.71}_{-0.71}$ \textbf{(2)} & 49.9\\
        2MASSI J1553022+153236& T7 & 15.34$^{+0.03}_{-0.03}$ & 15.76$^{+0.03}_{-0.03}$ & 15.94$^{+0.03}_{-0.03}$ & 13.32$^{+0.16}_{-0.16}$ \textbf{(2)} & 140.6\\
        SDSS J162838.77+230821.1  & T7 & 16.25$^{+0.03}_{-0.03}$  & 16.72$^{+0.03}_{-0.03}$ & 16.63$^{+0.03}_{-0.03}$ & 13.32$^{+0.16}_{-0.16}$ \textbf{(2)} & 42.8\\
        WISEPA J165311.05+444423.9 & T8 & 17.07$^{+0.02}_{-0.02}$ & 17.59$^{+0.05}_{-0.05}$ & 17.05$^{+0.07}_{-0.07}$ & 13.21$^{+0.33}_{-0.33}$ \textbf{(3)} & 38.7\\
        WISEPA J171104.60+350036.8PRZ0.5& T8 & 17.63$^{+0.13}_{-0.13}$ & 18.06$^{+0.14}_{-0.14}$ & 18.09$^{+0.14}_{-0.14}$ & 24.81$^{+1.48}_{-1.48}$ \textbf{(3)} & 32.1\\
        WISEPA J174124.26+255319.5& T9 & 16.18 $^{+0.02}_{-0.02}$ & 16.31$^{+0.04}_{-0.04}$ & 17.02$^{+0.2}_{-0.2}$ & 4.67$^{+0.06}_{-0.06}$ \textbf{(3)} & 122.5\\
        WISE J181329.40+283533.3& T8 & 16.92$^{+0.02}_{-0.02}$ & 17.11$^{+0.06}_{-0.06}$ & 16.92$^{+0.06}_{-0.06}$ & 13.59$^{+0.37}_{-0.37}$ \textbf{(1)} & 54.2\\
        WISEPA J185215.78+353716.3 & T7 & 16.33$^{+0.02}_{-0.02}$ & 16.72$^{+0.06}_{-0.06}$ & 16.5$^{+0.06}_{-0.06}$ & 15.06$^{+0.66}_{-0.66}$ & 125.9\\
        WISEPA J195905.66-333833.7& T8 & 16.71$^{+0.07}_{-0.07}$ & 17.18$^{+0.05}_{-0.05}$ & 16.93$^{+0.09}_{-0.09}$ & 11.72$^{+0.3}_{-0.3}$ \textbf{(3)} & 60.2\\
        WISE J200050.19+362950.1 & T8 & 15.44$^{+0.01}_{-0.01}$& 16.13$^{+0.04}_{-0.04}$  & 15.85$^{+0.01}_{-0.01}$ & \phn7.62$^{+0.17}_{-0.17}$ \textbf{(1)} & 67.3\\
        WISEPC J215751.38+265931.4& T7 & 17.05$^{+0.02}_{-0.02}$ & 17.49$^{+0.04}_{-0.04}$ & 17.34$^{+0.06}_{-0.06}$ & 15.92$^{+0.56}_{-0.56}$ \textbf{(1)} & 87.0\\
        WISEPC J220922.10-273439.5 & T7 & 16.6$^{+0.02}_{-0.02}$ & 16.95$^{0.06+}_{-0.06}$ & 17.35$^{+0.06}_{-0.06}$ & 13.81$^{+0.72}_{-0.72}$ \textbf{(1)} & 75.2\\
        WISEPC J221354.69+091139.4 & T7 & 16.77$^{+0.02}_{-0.02}$  & 17.12$^{+0.06}_{-0.06}$ & 17.11$^{+0.06}_{-0.06}$& 19.19$^{+1.14}_{-1.14}$ \textbf{(1)} & 77.0\\
        WISEPC J222623.05+044003.9 & T8 & 16.90$^{+0.02}_{-0.02}$  & 17.24$^{+0.09}_{-0.09}$ & 17.45$^{+0.07}_{-0.07}$& 18.38$^{+1.99}_{-1.99}$ \textbf{(3)} & 69.7\\
        WISEPC J225540.74-311841.8& T8 & 17.33$^{+0.01}_{-0.01}$ & 17.66$^{+0.03}_{-0.03}$ & 17.42$^{+0.05}_{-0.05}$ & 14.14$^{+0.84}_{-0.84}$ \textbf{(3)} & 50.6\\
        WISEPC J231939.13-184404.3 & T7.5 & 17.56$^{+0.02}_{-0.02}$ & 17.95$^{+0.05}_{-0.05}$ & 18.26$^{+0.08}_{-0.08}$ & 11.75$^{+0.43}_{-0.43}$ \textbf{(1)} & 31.3\\
        ULAS J232123.79+135454.9 & T7.5 & 16.72$^{+0.03}_{-0.03}$ & 17.15$^{+0.03}_{-0.03}$ & 17.16$^{+0.01}_{-0.01}$ & 11.96$^{+0.34}_{-0.34}$ \textbf{(3)} & 51.9\\
        WISEPC J234026.62-074507.2 &  T7 & 16.08$^{+0.03}_{-0.03}$ & 16.4$^{+0.03}_{-0.03}$ & 16.51$^{+0.06}_{-0.06}$ & 20.92$^{+1.36}_{-1.36}$ \textbf{(3)} & 95.9\\
        WISEPC J234841.10-102844.4 & T7 & 16.63$^{+0.02}_{-0.02}$ & 16.99$^{+0.06}_{-0.06}$ & 16.84$^{+0.06}_{-0.06}$ & 14.79$^{+0.83}_{-0.83}$ \textbf{(1)} & 71.3\\
        \bottomrule
    \end{tabular}
    
        
    \tablerefs{References for both spectral types and parallax distances in this table are (1) \cite{2020AJ....159..257B}, (2) \cite{2012ApJS..201...19D}, (3) \cite{2019ApJS..240...19K}, (4) \cite{2018A&A...616A...1G}, (5) \cite{2003AJ....126..975T}, (6) \cite{2012ApJ...752...56F}, (7) \cite{2008ApJ...689L..53B}, (8) \cite{2012ApJ...752...56F}, (9) \cite{Leggett_2012}}
    \end{threeparttable}
    \label{tab:photometry}
\end{table*}

Our entire dataset is derived from late-T dwarf IRTF SpeX prism observations (0.95 - 2.5 $\mu$m, R$\sim$120) largely described in \citep[see Table 3]{2021ApJ...916...53Z, 2021arXiv210505256Z}.  There are a total of 55 objects with SpeX spectra with 39 available within the SpeX Prism Library, and 16 from our own observational campaigns. Briefly, the sample is comprised of objects from the volume-limited ($d<$25 pc) survey in \cite{2020AJ....159..257B} with our SpeX objects out to 20 pc. 54 of the 55 have well measured parallaxes and all spectra are flux calibrated using MKO H-band photometry. Our analysis, for uniformity purposes, also includes the 11 objects from \cite{2017ApJ...848...83L}. Due to low signal-to-noise, our final analysis excludes 5 objects, for a total of 50 in our analysis. This subset is highlighted in a color-magnitude diagram shown in Figure \ref{fig:color} with photometry and parallax measurements enumerated in Table \ref{tab:photometry}. As in our past works we only analyze every 3rd spectral point as SpeX samples the line-spread shape with $\sim$2.5 pixels.

\section{Methods} \label{sec:Methods}

Our retrieval framework follows closely that described in \cite{2017ApJ...848...83L} and \cite{2019ApJ...877...24Z}. The model computes the thermal emission spectrum of a brown dwarf and includes 31 free-parameters to describe the atmosphere: constant-with-altitude volume mixing ratios (VMR's) for: H$_2$O, CH$_4$, NH$_3$, CO, CO$_2$, H$_2$S, Na, and K (8 in total), gravity, radius-to-distance scale factor $(R/D)^2$, 15 independent temperature-pressure (TP) profile points implemented within a Gaussian-Process-like smoothing framework, and a simple cloud parameterization \citep{2006ApJ...650.1140B}; summarized in Table \ref{tab:parameters}.

Compared to our previous work we have made several adjustments to this model. We have separated the alkali constraints into distinct Na and K free parameters (as opposed to Na+K fixed at a solar ratio), removed H$_2$S due to low predicted abundances at these temperatures, and updated the alkali metal wing profiles \citep{2016A&A...589A..21A}. All other opacity sources are identical to \cite{2017ApJ...848...83L} and \cite{2019ApJ...877...24Z}. Model cross-section sampling is used at a constant R=10000, a resolution proven sufficient for interpreting SpeX brown dwarf observations \citep{2015ApJ...807..183L}. Parameter prior ranges are the same as those of \cite{2015ApJ...807..183L} which are restated in Table \ref{tab:priors}. Fits were performed as in the mentioned previous studies using the {\tt emcee} package \citep{2013PASP..125..306F} with 224 walkers run out to 60,000 iterations.  To assure convergence several test cases were ran to between 120K and 1 million iterations with no significant differences in model posteriors. {\tt emcee}, like all MCMC-based methods, requires an initial guess to initiate the walkers. The initial guess is based on a ``Gaussian-ball" about a lose by-eye fit to the spectra of each object. The final solutions were found insensitive to the initial guess as was the case in previous publications.

To reduce computational runtime for such a large number of objects, our radiative transfer core (Appendix A in \cite{1991JGR....96.9027L}) was rewritten to make use of graphics processing units (GPUs). This was done using the Anaconda Numba {\tt guvectorize} framework on NVIDIA Tesla V100 GPUs. Forward model times improved by a factor of $\sim$100 (0.01s or so per model, at an R=10,000 over the 0.95 - 2.5$\mu m$ wavelength range). Given the limited memory of the GPUs (32 GB), we could only run up to 16 simultaneous CPU threads at a time. The overall computational improvement between this work and that of \cite{2017ApJ...848...83L} is about a factor of 10 (about 6 hours to hit 60,000 iterations). GPU and CPU routines were tested to produce identical model spectra with no impact on any science results.

\begin{table}[b]
    \caption{Retrieved Parameters}
    \label{parameters}
    \addtolength{\leftskip} {-0.8cm}
    \begin{tabular}{c p{6cm}}
        \toprule
        Parameter & Description\\
        \midrule
        log (f$_{i}$) & log of the Volume Mixing Ratio (VMR) of a gas species that is constant with altitude. Gases that are considered include H$_2$O, CH$_4$, CO, CO$_2$, NH$_3$, K, and Na\\
        log (g) & log of surface gravity [cm s$^2$]\\
        (R/D)$^2$ & Radius-to-distance scale factor [$R_{\rm Jup}$/pc]\\
        T$_{i}$ & Temperature (in Kelvin) at a given pressure level\\
        ${\Delta}{\lambda}$ & Wavelength calibration uncertainty [nm] \\
        ${b}$ & errorbar inflation exponent\\
        $\gamma,\beta$ & TP-profile smoothing hyperparameters (eq. 5, \cite{2015ApJ...807..183L})\\
        ${\kappa}_{P_0}$, ${P}_{0}$, ${\alpha}$ & Cloud opacity profile parameters (eq. 1, \citep{2017ApJ...848...83L}, cloud base opacity, cloud base pressure, cloud fractional scale height)\\
        \bottomrule
    \end{tabular}
    \label{tab:parameters}
\end{table}

\begin{table}[b]
    \caption{Parameter Priors}
    \centering
    \footnotesize
    \begin{tabular}{c c}
        \toprule
        Parameter & Prior\\
        \midrule
        log (f$_{i}$) & $\textgreater$ -6, $\Sigma$f$_{i}$=1\\
        log (g), (R/D)$^2$ & M $\textless$ 80M$_{Jup}$\\
        ${\Delta}{\lambda}$ & (-10 , 10)\\
        10$^{b}$ & (0.01$\times$min($\sigma_i^2$) , 100$\times$max($\sigma_i^2$))\\
        $\gamma$ & Inverse Gamma($\Gamma$($\gamma$; $\alpha$, $\beta$)), $\alpha$=1, $\beta$= 5 $\times 10^{-5}$\\
        log($\kappa_{C}$), log(${P}_{0}$), ${\alpha}$ & (-12, 0), (2.3, 2.8), [0,10)\\
        \bottomrule
    \end{tabular}
    \label{tab:priors}
\end{table}



\section{Results \& Discussion} \label{sec: results} 
 As it is largely uninformative to show the full posteriors and discuss each individual object (available in the attached Zenodo link\footnote{\href{https://doi.org/10.5281/zenodo.6578764}{https://doi.org/10.5281/zenodo.6578764}}), here we simply summarize the key results broken down by thermal structure, composition (molecular abundance trends/chemistry and elemental abundances), and evolutionary parameters.
 
\begin{figure*}
    \centering
    \includegraphics[angle=0, width=17cm, ]{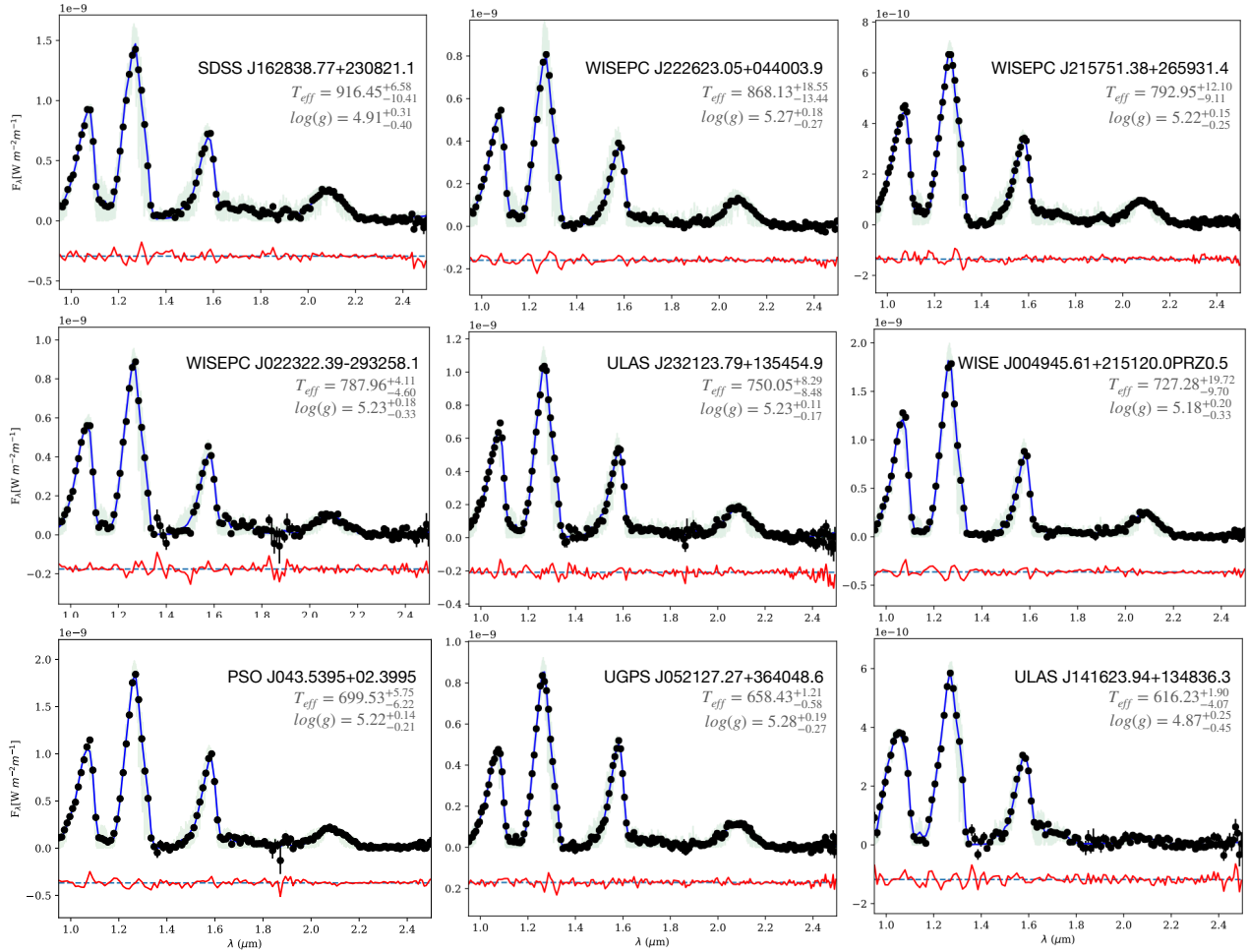}
    \caption{Subset of the data (black round points with errors) and resulting fit (binned best fit in blue, residuals about the dashed line below in red). In blue is a higher resolution (R$\sim$25,000) representation of the best fit spectrum. The effective temperature and gravity constraints are given in the upper right hand corner of each panel.}
    \label{fig:spectra}
\end{figure*}

 Figures \ref{fig:spectra} and \ref{fig:tp_profiles} provide a snapshot of the fits and temperature-pressure profiles, respectively, of a sub-sample of representative objects. The best fits and temperature profiles for all objects in our sample are provided in the Appendix (Figures \ref{fig:50_spec} and \ref{fig:50_tp-profiles}). As has been discussed in previous papers in this series, the visual quality of the model fits in Figures \ref{fig:spectra} and \ref{fig:50_spec} are substantially better than those by more traditional grid-based models, with residuals following only the uncertainty in the data itself with little other deviation. 
 
 Table \ref{tab:retrieved_and_derived} summarizes the nominal constraints for the key properties of individual objects. As in our past works, the effective temperatures are derived by integrating over an ensemble of best fits for each object, extrapolated out to between 0.7 - 20 $\mu m$, and radius is derived from the retrieved $(R/D)^2$ scale factor and the measured distances (from Table \ref{tab:photometry}).  The elemental C and O abundances are derived from the retrieved molecular abundances (more details below).  When comparing the retrieved quantities to those predicted from self-consistent grids (given an effective temperature and gravity and assuming solar abundance chemistry and cloud free), we refer to those produced by the ScCHIMERA model \citet{Piskorz_2018, 2018ApJ...855L..30A, 2018AJ....156...10M, 2018ApJ...858L...6K, 2019ApJ...872...27G, 2019ApJ...877...24Z, 2020arXiv200505153C, 2020arXiv200610292B, 2021arXiv210601387M}, with additional eddy-mixing for the NH$_3$-N$_2$ and CO-CH$_4$ chemistry \citep{2014ApJ...797...41Z}.
 
\subsection{Thermal Structures} \label{sec:Thermals}

Late-T type brown dwarfs are ideal for temperature profile retrievals due to the lack of obscuring, optically thick condensate clouds that would otherwise limit the range of pressures that could be probed via spectroscopy. The high degree of flexibility of the thermal profile parameterization allows us to compare thermal profiles derived directly from observational spectra against assumptions often made in more self-consistent models. Since the thermal structure is a reflection of the total energy balance, identifying similarities and differences between retrieved profiles and those predicted from 1D radiative-convective-thermochemical equilibrium (RCE) for such a large number of objects enables us to assess and quantify potential sources of heating outside of the standard RCE assumptions.

Figure \ref{fig:tp_profiles} summarizes (see Appendix, Figure \ref{fig:50_tp-profiles} for all objects in the sample) the retrieved thermal structures for nine representative objects (red) compared with 1D RCE predicted TP profiles (blue). Uncertainties for the RCE temperature profiles are derived by propagating uncertainties in the retrieved log($g$), \teff, metallicity and C/O. Overlaid for context are condensation curves for several notable condensate species \citep{2012ApJ...756..172M, 2014ApJ...787...78M}. As in previous studies we reiterate that our SpeX observations largely probe only the region between 0.1 and 10bars with the rest of the profile's constraints being largely dependent upon the smoothing prior; any structure outside this range should be interpreted with caution. 

We find that the uncertainties on the retrieved profiles are similar to previous studies, with 1$\sigma$ errors on the order of 100K with larger deviations being owed to larger uncertainties on the spectra themselves \citep{2015ApJ...807..183L, 2017ApJ...848...83L, 2019ApJ...877...24Z}. While some objects appear to be in agreement with the assumption of 1D RCE to 2$\sigma$ (see WISEPC J0223 in Figure \ref{fig:tp_profiles}, middle left) 18 objects are discrepant with the 1D RCE predicted profiles. There is a systematic offset whereby the retrieved profiles for a given \teff and log(g) are hotter (at a given pressure) than the corresponding 1D RCE predicted profiles. In addition to this systematic offset, 9 objects display an interesting ''kink" in the profiles near the 10bar pressure level.

We first tested if the overall offset was physical or a simple result of over or under-constraining either model. To make clear, the error bounds for the grid model are calculated by taking the constraints from the retrieval model and propagating them into error bounds for the grid's thermal profile using several key parameters: \teff, log(g), metallicity, C/O, as well as a reasonable range of K$_{zz}$ values for late-T dwarfs. We tested the robustness of these constraints by artificially inflating the errors of the retrieval results by 1$\sigma$ to see if the resulting thermal profiles would then overlap. Inflating the errors in this way removed the evidence of any kink structure near 10bars. As this pressure range is near the edge of the contribution functions that our spectra probe, the presence of such a feature should be interpreted with caution.


The most likely reason for the retrieval profiles being systematically hotter than the 1D RCE profiles is owed to how \teff is computed in the retrieval model coupled with the grid-based models being forced onto RCE. As explained in previous papers, a sample of several thousand spectra are taken from the posterior and propagated out to roughly 20$\mu$m. This flux is then integrated and \teff is computed from L$_{Bol}$. However, its important to note that a majority of the output flux for these T dwarfs lie at longer wavelengths outside the range of our SpeX dataset. Since we cannot detect molecules such as CO or CO$_2$ in our SpeX data, these major absorptive species at longer wavelengths may not be properly included in these models which stretch out to 20$\mu$m. Therefore the result may be an over-estimate for the flux and thereby \teff. This estimate for \teff is then used as an input into the 1D RCE model which, while at similar \teff, may be limited by being forced onto a strict 1D RCE prescription for the PT profile.  Another complicating factor is that, as shown in Figure \ref{fig:tp_highest_lowest_logg} and discussed in Section \ref{sec:metlogg}, some objects can be offset by several hundred kelvin depending upon the choice for priors on log(g).

If we assume the offset is physically motivated, then it might be possible that these late-T dwarfs may be impacted by the presence of iron-silicate condensate clouds. To test this we calculated the cloud column optical depth, $\tau_{cld}$ by integrating the random draws of our retrieved opacity profile from our cloud model. Figure \ref{fig:tau_cloud} shows the result for two representative objects (all targets available in the above Zenodo link), where the most likely values of $\tau_{cld}$ are far below unity. This suggests that optically thick clouds do not strongly affect our dataset. Even targets such as ROSS 458C, which has been seen in previous works to better fit grid-based models that incorporate these clouds \citep{2010ApJ...725.1405B,2011MNRAS.414.3590B,2012ApJ...756..172M}, still strongly favors a cloud-free retrieval result.

Additionally, these targets may have additional heating mechanisms in their atmospheres that could be driving the system away from pure radiative-convective equilibrium. However it seems more likely that differences in the physical and chemical assumptions between the 1D RCE models and retrieval can bias the thermal profile using the same \teff.

\begin{figure*}[t]
    \centering
    \includegraphics[angle=0, width=18cm]{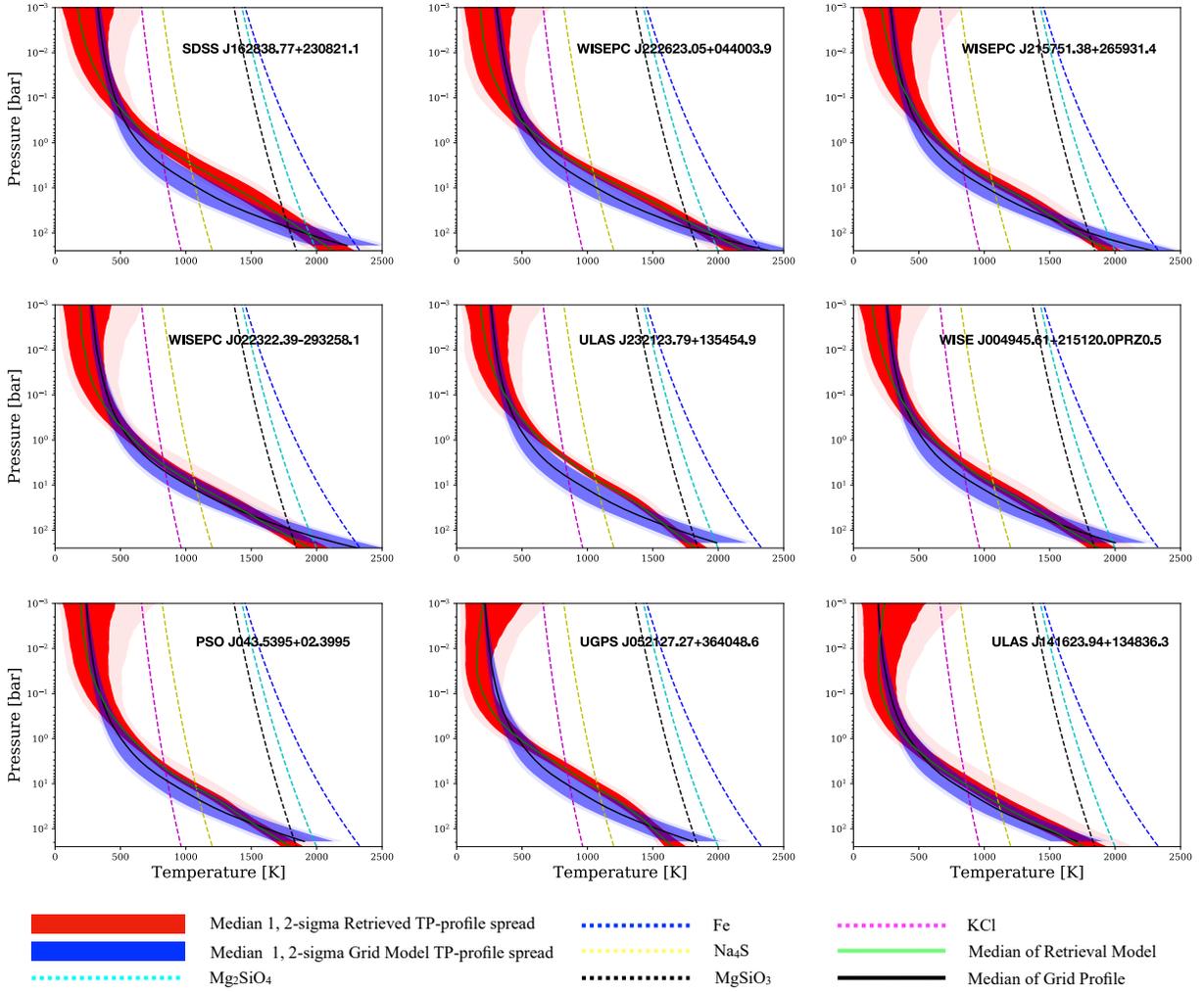}
    \caption{TP-profiles of representative T dwarfs of our collection of 50 objects. The thermal profiles indicated here are of the retrieved objects (in red), and of grid models (in blue, see text) that they closely compared to based on their similarities in a mix of \teff, log(g), C/O, and metallicity parameters. For each of the thermals profiles, both 1 and 2$\sigma$ confidence intervals are shown. Equilibrium condensation curves of Mg$_2$SiO$_4$, Fe, MaSiO$_3$, Na$_2$S, and K are shown by the cyan, blue, black, yellow, and magenta dashed lines, respectively.}
    \label{fig:tp_profiles}
\end{figure*}

\subsection{Atmospheric Composition} \label{sec: Atmospheric composition}
\subsubsection{Molecular Abundances} \label{sec: molecular abundances}
\begin{figure*}[t]
    \centering
    \includegraphics[width=1.0\textwidth]{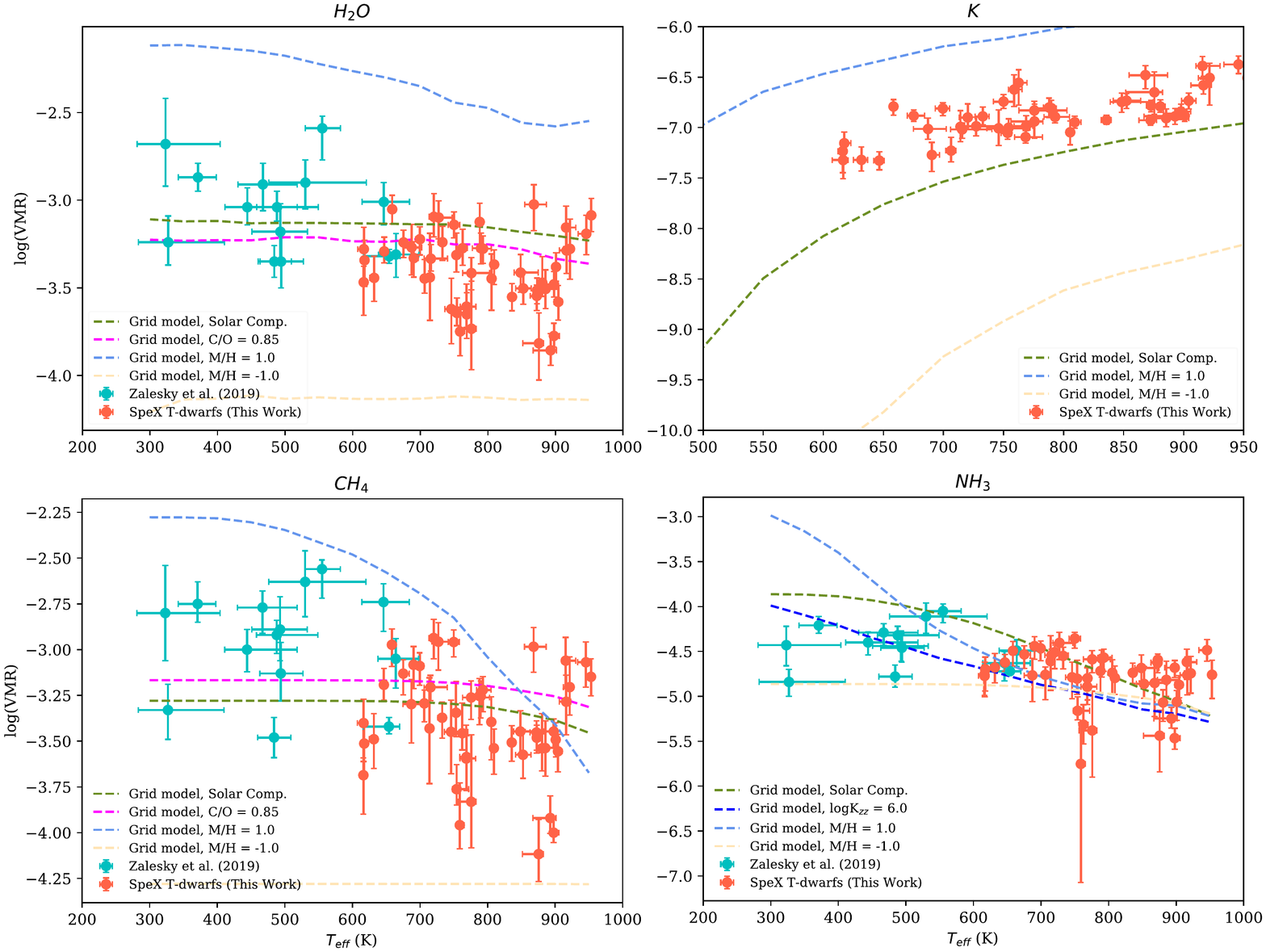}
    \caption{Molecular abundances (expressed as log of the VMRs) of H$_2$O, CH$_4$, NH$_3$, and K plotted against \teff\ for T dwarfs from this study (color:tomato) and brown dwarfs from \cite{2019ApJ...877...24Z}. For K, objects from \cite{2019ApJ...877...24Z} are not included because that work considered Na+K as one alkali species, whereas herein we are treating them as individual species. The four sub-plots show how the molecular and alkali abundances of each species varies with \teff.  All the modelled molecular abundance trends are defined by log(g) = 5.0 and solar composition (M/H = 0.0, C/O = 0.5), and we assume equilibrium chemistry unless otherwise specified. We observe that the overall trends of water, methane, and ammonia are consistent with solar chemistry curve models.
    }
    \label{fig:mol_abundance}
\end{figure*}

A key goal of this investigation is to develop a large, diagnostic dataset of retrieved molecular abundances across the late-T spectral class to serve as a baseline unirradiated comparison for directly imaged and transiting exoplanets. For these objects, species such as H$_2$O, CH$_4$, NH$_3$, and K are the most readily accessible over the SpeX wavelength range. The change in the abundances of these gases with \teff are diagnostic of the driving physical and chemical processes in brown dwarf atmospheres \citep{1999ApJ...512..843B, 2002Icar..155..393L, 2007ApJS..168..140S}.

Figure \ref{fig:mol_abundance} shows the behavior of the retrieved molecular abundances with temperature combined with the retrieved Y-dwarf molecular abundances from \cite{2019ApJ...877...24Z} (cyan). Overlaid for context are predictions from the 1D RCE models over a range of atmospheric metallicity and C/O ratios. Unless labeled otherwise each curve assumes a solar composition ([M/H]=0, C/O=0.55). To compute effective atmospheric abundances from the 1D RCE models, gas volume-mixing-ratios (VMRs) are averaged over the 1 - 30bar pressure levels based on the range of pressure levels probed by the SpeX spectra. The exact retrieved abundances for all 50 objects are enumerated in the Appendix under Table \ref{tab:retrieved_and_derived}. We find that the precision of our constraints is slightly improved relative to \cite{2019ApJ...877...24Z} as expected from the increase in quality of data. For all species we find that our abundances are constrained to within $\sim$0.25dex at 1$\sigma$. 

For both H$_2$O (top left) and CH$_4$ (lower left) there is a trend of decreasing abundance with higher \teff, consistent with what was found in \cite{2019ApJ...877...24Z}. We are able to confirm that this trend extends beyond two late-T dwarfs and continues with increasing temperature though there is considerable spread. 
For H$_2$O, this spread is consistent with what is expected for these objects between a metallicity of 0.1 to 10 times solar and various C/O ratios as shown. As there is no known physical mechanism for why these two populations should be disparate in their water content, and given the error bounds and sampling size of the Y-dwarfs, it seems likely that such a trend is a result of low-sampling from this population. Future studies which could retrieve higher precision ($\sim$0.25dex) water abundances for brown dwarfs between 300-700K would strongly help either confirm or deny such a trend with \teff.

For CH$_4$, most objects are within the predicted ranges of from the grid model. However, there are interestingly several objects with \teff between 900-1000K which deviate from this assumption. This would require either anonymously high C/O ratios (some well-above 1.0), high metallicities, or both when compared to the curves from the grid model. These objects must be interpreted with caution for two reasons. First is that only 4 objects lie above the expectations from the grid model at only 1$\sigma$. Second, we emphasize such curves do not represent all possible combinations of parameter space. Despite this, there may be physical motivation for finding objects with anomalously high CH$_4$ abundances near 900K. It has been well studied that L and T dwarfs have both theoretical and observational evidence for the presence of CO in their photospheres driven by disequilibrium convective mixing \citep{1997ApJ...489L..87N, 1998ApJ...502..932O, 2003IAUS..211..345S, 2004AJ....127.3516G, 2006ApJS..166..585B, 2009ApJ...695..844G, 2012ApJ...760..151S, 2012ApJ...748...74L}. More recently, \citet{2020AJ....160...63M} found that the CO constraints for several brown dwarfs between 250-700K suggest that at lower temperatures ($\sim$250K) there is evidence for stronger mixing (higher CO, less CH$_4$) and at higher temperatures ($\sim$700K) there is evidence for reduced mixing strength (less CO, more CH$_4$). It is possible that we could be seeing objects that follow such a trend up to higher temperatures, however given the low number of objects in our sample and the precision of our constraints, this is largely inconclusive. There is a wealth of available early T and L spectral class objects to study to confirm such a trend, but have the added complication of obscuring condensate clouds which is beyond the scope of this work.

We find that the retrieved NH$_3$ abundances are largely consistent with grid model expectations, with the exception of 9 objects which show additional NH$_3$ near 900K. This result is surprising given that the onset of ammonia is the defining feature of the Y-dwarf spectral class. While again this could be a result of small number sampling, it is possible that this could be evidence for enhanced mixing strength at higher temperatures resulting in an excess of NH$_3$ compared to expectations from equilibrium. While \citet{2020AJ....160...63M} found no evidence for enhanced mixing at these temperatures, disequilibrium abundances of NH$_3$ have been detected in the past \citep{2007ApJ...656.1136S, 2009ApJ...695..844G, 2010ApJ...710.1627L,2015ApJ...799...37L}.

As was found in Part II and III, there is a strong systematic trend of decreasing potassium between the late-T to early-Y spectral types as a result of equilibrium rainout chemistry removing aluminium and silicate reserves near 1300K, leading to a delayed depletion of Na+K closer to \teff $\sim$700K where they condense into KCl and Na$_2$S. Figure \ref{fig:mol_abundance} shows the the trend with our larger sample which continues to indicate a steadly decreasing trend with decreasing \teff, with potassium abundances falling roughly 1 dex between -6.5 and -7.5 log(VMR) consistent with what was found in Parts II and III. We do not show the results from the Y dwarfs here as in that work, Na and K were treated as a single gas with most objects only providing upper limits.

While we are not able to fully explain the observed abundance trends with standard 1D RCE assumptions, it may indicate that some underlying physical or chemical process is not being taken into account in these models. Regardless, this sample of retrieved abundances provides a unique dataset to further explore the chemistry of sub-stellar atmospheres.

\begin{figure}[b]
    \addtolength{\leftskip}{-1.5cm}
    \includegraphics[width=1.4\columnwidth]{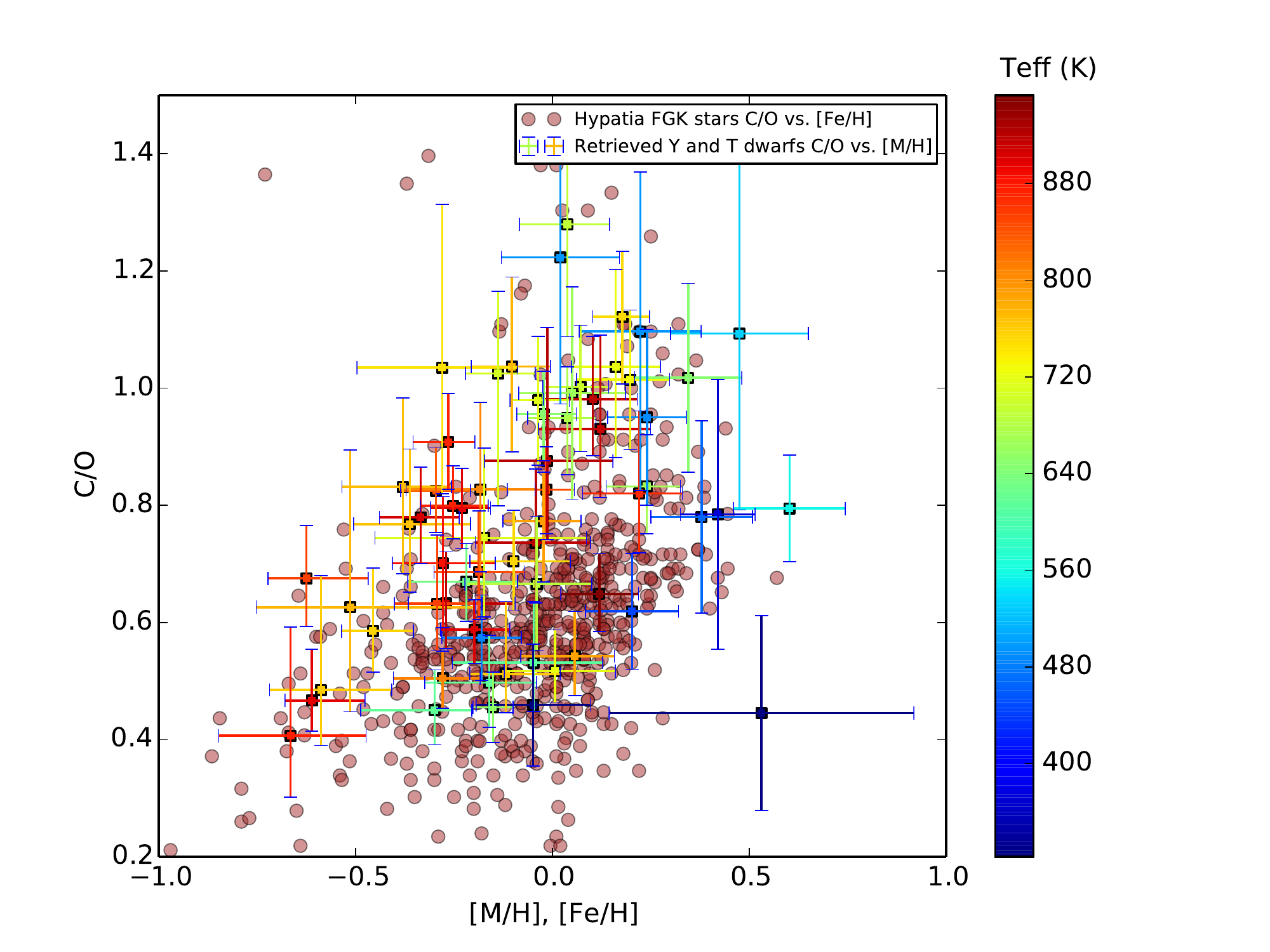}
    \caption{C/O vs. bulk metallicity of our Y and T dwarfs ensemble. The brown dwarfs are color-coded according to their effective temperatures. For comparison with field stars, we overplot the brown dwarfs (C/O vs. [M/H]) with Hypatia FGK stars(C/O vs. [Fe/H]) that are within comparable parallax distances ($\leq$ 30 pc) to that of the brown dwarfs.  The two populations have highly similar distributions owed to their similar formation environments.}
    \label{fig:metallicity}
\end{figure}

\begin{figure}[b]
    \addtolength{\leftskip}{-0.5cm}
    \includegraphics[width=1.2\columnwidth]{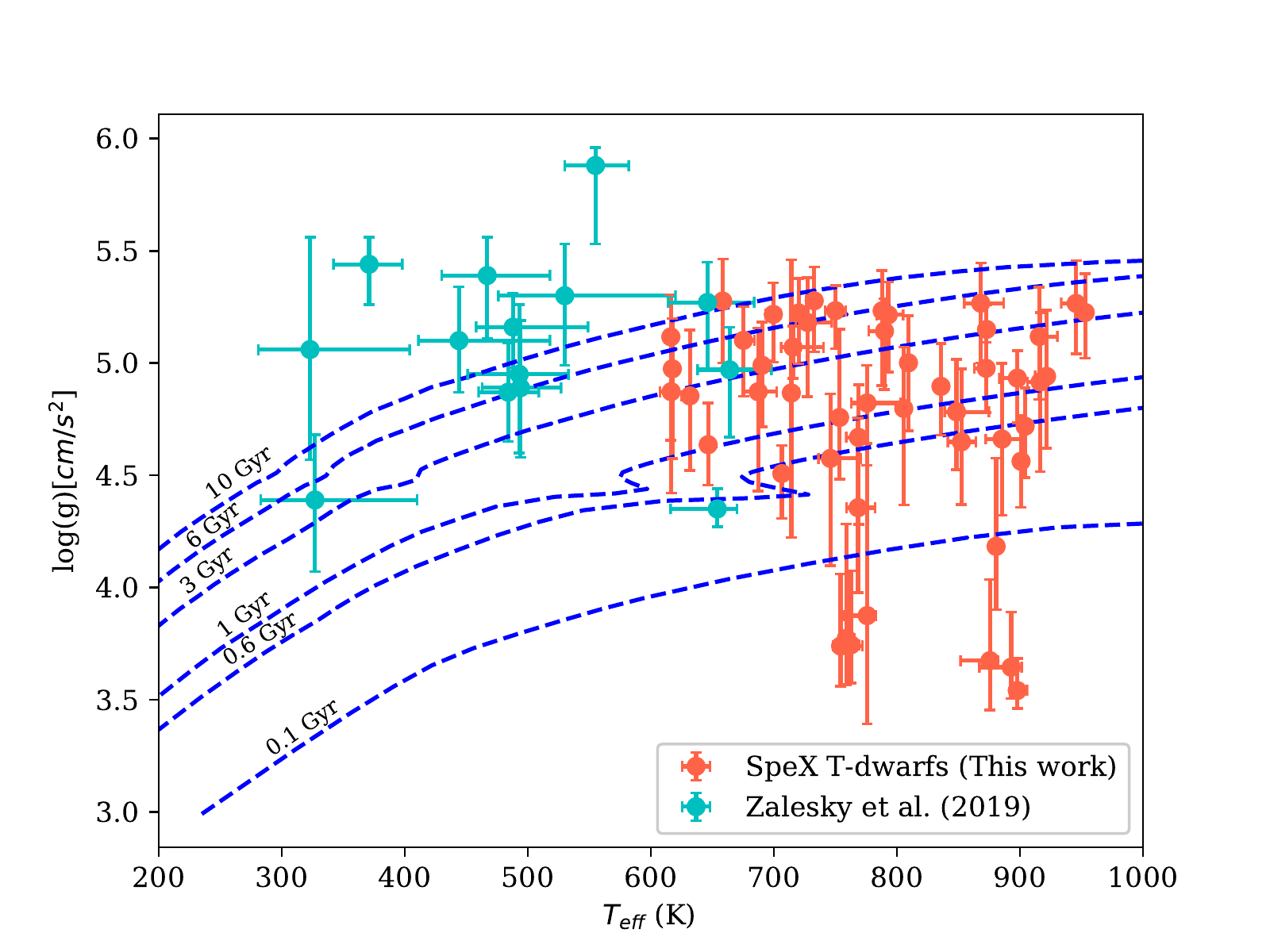}
    \caption{log g vs. \teff\ for our collection of early Y dwarfs and late-T dwarfs. Here we notice that we almost have a handful of T dwarfs with low surface gravity values while Y dwarfs tend to have higher gravity values. Based on evolutionary models from (Marley et al. (2020), in prep) (shown by blue dashed curves), T dwarfs having the lowest log g values would be relatively young ($\sim$100 Myr) while Y dwarfs having the highest log g would be relatively old ($\sim$12 Gyr). Most of the objects above the 10 Gyr curve are considered atypical although their high gravities can be justified by the logic that under-estimations of gravities that are done in grid models \citep{2015ApJ...804...92S, 2019ApJ...877...24Z}.}
    \label{fig:logg1}
\end{figure}

\subsubsection{Metallicity and Elemental Ratios} \label{sec:metallicity and C-O}

In addition to the molecular composition of a brown dwarf atmosphere, the total elemental abundance inventory, and ratios between those abundances, can assist in placing brown dwarfs in a larger context with their stellar and planetary-mass cousins. Stars and field brown dwarfs are expected to form via fragmentation and core-collapse of a molecular cloud. Thus one would expect their elemental inventories to be highly similar. In contrast, planets that are formed in disks might have a completely different elemental composition due to a wide variety of processing mechanisms, differential ice-lines and thus elemental ratios within the disk, and/or migration of the planet within the disk \citep[e.g.][]{2011ApJ...743L..16O, 2014Life....4..142H}. These mechanisms can change the elemental composition to range from anywhere from stellar composition \citep{2013ApJ...775...80F, 2016ApJ...832...41M}, to vastly different metal enrichment, and/or to have completely different C/O ratios, sometimes several factors greater or lesser than the host star \citep{2014ApJ...794L..12M, 2016A&A...595A..83E}. Identifying at what masses this dramatic compositional shift may occur for a large sample of brown dwarfs can better place brown dwarfs as a whole in a larger astronomical context. 

As noted in previous studies in this series, the bulk of the metals found in brown dwarf atmospheres with \teff$\leq$1000K are largely comprised of C and O contained in the water and methane content with a smaller amount of N stored in NH$_3$. The metallicity is computed as,

\begin{equation}
    [M/H] = log(\frac{(M/H)_{\rm T dwarf}}{(M/H)_{\rm solar}})
    \label{equ:norm_metals}
\end{equation}

where the metallicity of the T dwarf is taken to be the summation of the number of elemental species included in the retrieval model. The C/O ratio is taken as,

\begin{equation}
    \frac{C}{O} = \frac{\Sigma{C}}{\Sigma{O}} \sim \frac{CH_4 + CO + CO_2}{H_2O + CO + 2CO_2}.
    \label{equ:co}
\end{equation}



Figure \ref{fig:metallicity} shows the [M/H] compared against our computed C/O for both this study and those objects in \cite{2019ApJ...877...24Z}. For context a representative sample of local FGK stars are overlaid (dark red) \citep{2014AJ....148...54H}. As in Parts II and III, we note that we additionally weight our water abundance by a factor of 1.3 to account of the loss of oxygen to condensate rainout species such as enstatite and forsterite \citep{1994Icar..110..117F}. We find that, in agreement with previous papers in this series, the broad-scale metallicities and C/O ratios for our brown dwarfs largely match both in value and in distribution that of the local FGK stellar population. The new population (largely green-orange color) seems to fill in the gap between our two previous late-T and early-Y populations. Additionally, we performed a two-dimensional, two-sample, Kolmogorov–Smirnov test and found a p-value of 0.18 between the brown dwarf and stellar populations \citep{PresTeukVettFlan92}. This suggests that our measurements display no statistically significant difference from the local FGK population. Though there is considerable scatter, there is a slight positive correlation between metallicity and \teff.

\subsection{\teff \& Gravity} \label{sec: Basics physical properties}

A brown dwarf's effective temperature and gravity are highly diagnostic of both its age and evolutionary history \citep{2001RvMP...73..719B, 2003A&A...402..701B, 2008ApJ...689.1327S}. Given that our sample should be comprised of largely field-age brown dwarfs, deviations in the these parameters from evolutionary models serves as both a sanity-check for our retrieval model, but may also serve as a sign of youth if a particularly low surface gravity is found.

Figure \ref{fig:logg1} shows our sample of late-T objects (orange) against results from \cite{2019ApJ...877...24Z} (cyan). Over-plotted are evolutionary cooling curves for several different ages (blue-dashed, Marley et al. 2021, in prep). A large and extensive discussion for why the population from \cite{2019ApJ...877...24Z} have anomalously high gravities is given in that paper. 

Here we focus on our sample, which for a roughly forty objects agree with being near 1-3Gyr at 1$\sigma$. However, seven T dwarfs appear to have lower surface gravity values compared to the rest of the objects by several sigma with ages of near 0.1Gyr. Additionally several objects are within 1$\sigma$ of being consistent with a 10Gyr age. Both of these ages are unusual given our sample and necessitate a sanity check of our retrieval model.

\begin{figure*}[t]
    \centering
    \includegraphics[width=1.0\textwidth]{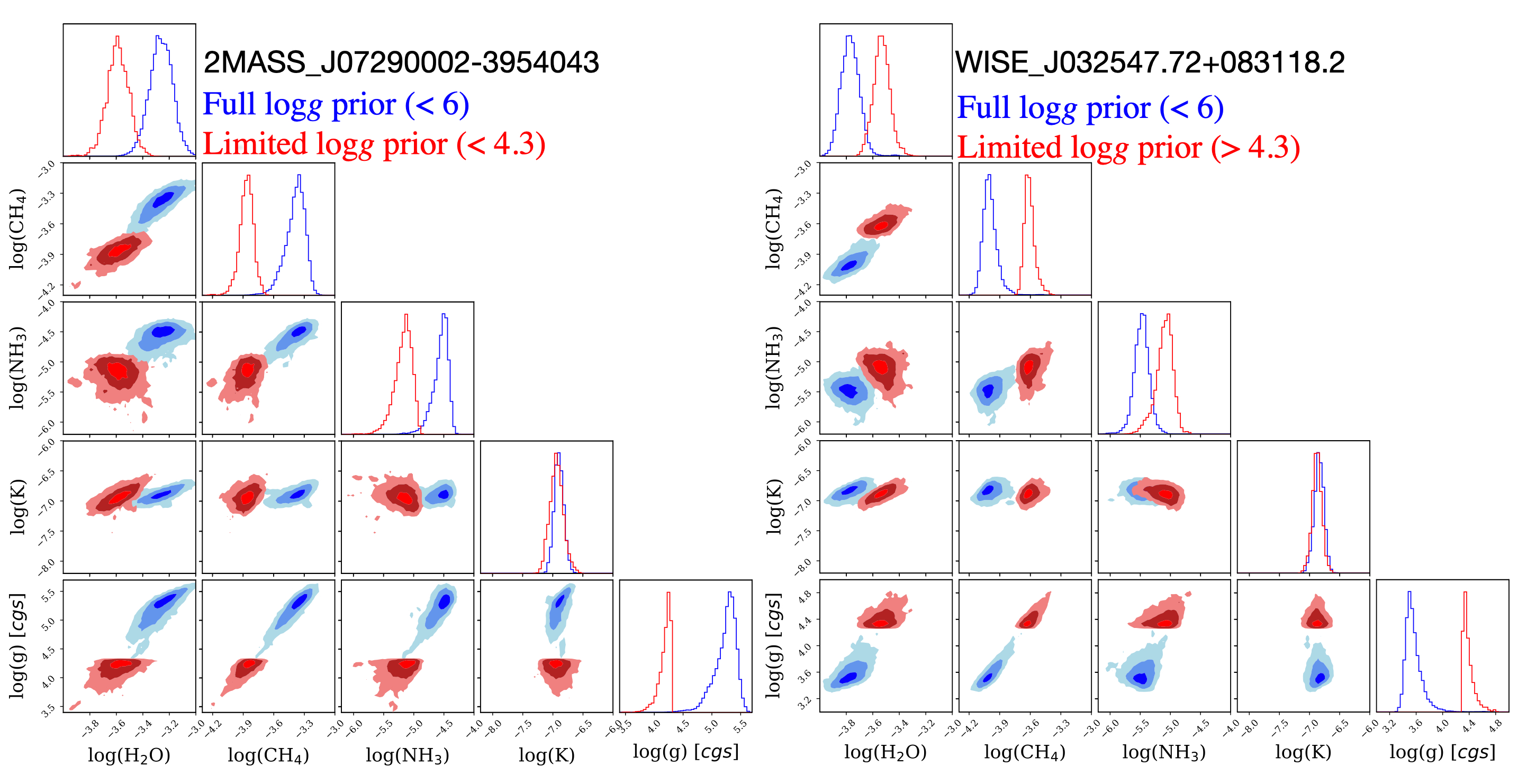}
    \caption{Corner plot posteriors of 2MASS-J072290002 (highest gravity object) and WISE-J032547.72 (lowest gravity object). This is based on a test of constraining priori ranges of gravity for the two objects. Gravity range restrictions affect gravity's correlation with the other gas species. Limited logg priors' results for surface gravity lean more toward full log(g) priors' results.}
    \label{fig:stairs_logg_expt}
\end{figure*}

The seven youngest objects are WISE J1254, PSO-J224, ROSS-458C, WISE-J1257, WISE-J1322, WISE-J1959, and WISE J0325. All the aforementioned T dwarfs, with the exception of ROSS-458C, have relatively low water and methane abundances. Furthermore, all seven of these T dwarfs have very low ammonia abundances. This observation of brown dwarfs exhibiting low gravities \textit{and} metallicites is highly atypical \citep{2014A&ARv..22...80H, 2011ApJ...732L..14Y, 2014A&ARv..22...80H}. To test the credibility of the surface gravity constraint, we conducted an experiment on both the lowest gravity object (WISE 0325) and on one of our two highest gravity objects (2MASS 0729) by varying their gravity priors.

\begin{figure*}
    \centering
    \includegraphics[width=0.9\textwidth]{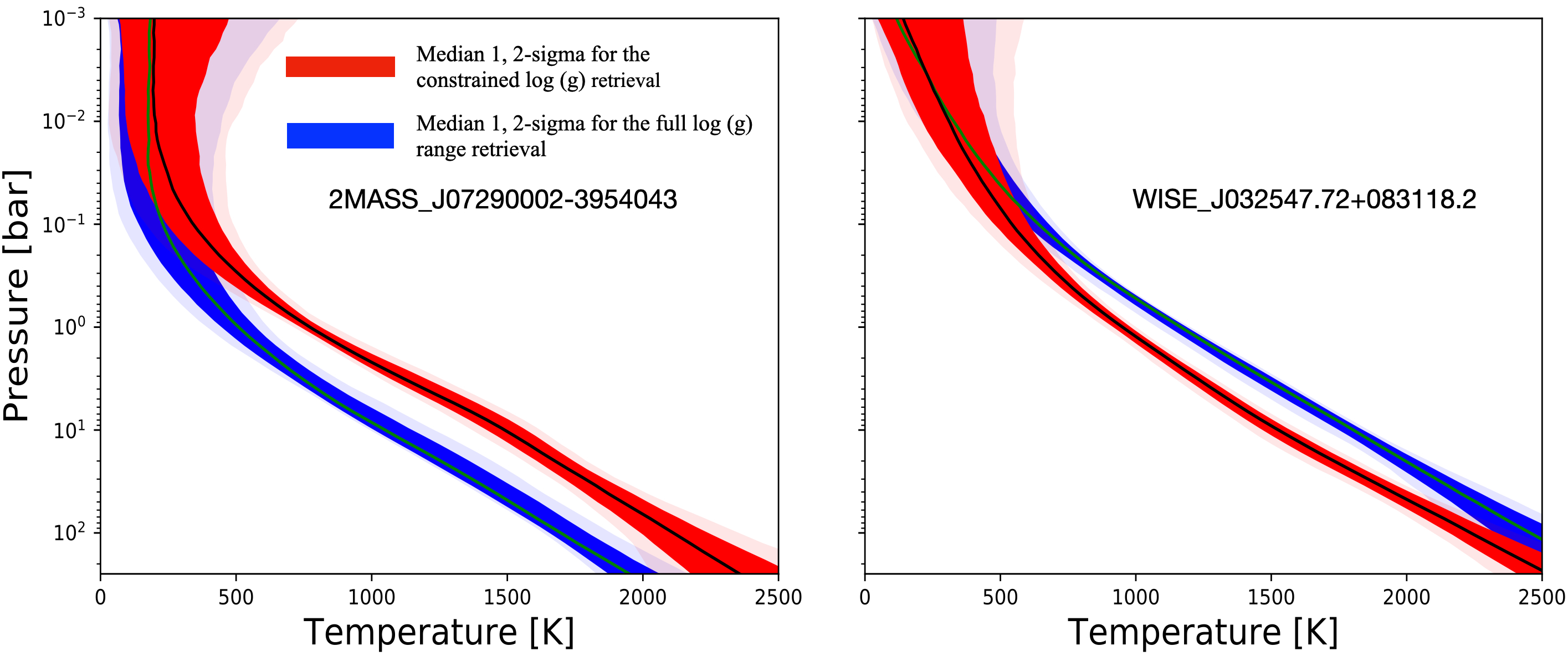}
    \caption{TP-profiles comparisons of limited and full logg ranges retrievals for 2MASS-J07290002 and WISE-J032547. Restrictions in gravity ranges influence the shape of thermal profiles.}
    \label{fig:tp_highest_lowest_logg}
\end{figure*}

\subsubsection{Metallicity - Surface Gravity Relationship} \label{sec:metlogg}

In this test we took a representative object that had both a relatively high and low log(g) constraint. The low-gravity object, WISE 0325, was used to perform a retrieval with a "full'' log(g) prior range of 0 $\leq$ log(g) $\leq$ 6 compared to a retrieval done with a "constrained'' log(g) prior range of 4.3 $\leq$ log(g) $\leq$ 6. The high-gravity object, 2MASS 0729, was set to 0 $\leq$ log(g) $\leq$ 4.3 while the full range was maintained at 0 $\leq$ log(g) $\leq$ 6. In the constrained cases, the gravity ranges were deliberately restricted to lower and higher gravity ranges to see how the retrieval results changed if forced to be in agreement with the system's likely true age. 

Figure \ref{fig:stairs_logg_expt} shows stair-pairs plots for both object's molecular abundance and gravity constraints in both the constrained (red) and unconstrained (blue) test cases. The constraint on log(g) shows a strong several-sigma offset between the two cases and importantly that the posterior in the constrained case is highly non-Gaussian as it is pushing up against the limit of the prior. This indicates that, despite being ''guided" to a lower/higher gravity in-line with field-age expectations, the spectrum itself has features which favor a higher/lower gravity than the priors allow. 

Alterations in gravity prior ranges also impact the posteriors of other species like water, ammonia, and methane as the parts of the spectra where gravity would be most constraining are the same regions where the absorption of these gases are most prominent. Potassium notably appears to be insignificantly affected by changes in gravity. This is likely because potassium absorption features only affect a small portion of the spectrum between 0.95-2.5 $\mu$m, mostly in the Y and J bandpasses. This high correlation between gravity and metallicity also impacts the constraints on the thermal profiles as well. as shown in Figure \ref{fig:tp_highest_lowest_logg}. 

Though a large majority of our objects are within reasonable expectations for field age brown dwarfs, the age estimates from several of these objects are highly nonphysical being either less than 0.1Gyr or older than the age of the Universe in the most extreme cases. These results show that retrievals, 1D RCE models, and comparison/sanity checks between both are needed to fully understand the atmospheres of brown dwarfs. One area this is possible is with different types of benchmark systems, where either metallicity or age is constrained. In our sample we have three systems which are in a binary (ROSS 458B,HD 3651B) or quaternary (Gliese 570D), all of which are within 1$\sigma$ for their stellar companion's age and metallicity, giving confidence to the reliability of our retrieval results despite the many caveats of using this technique.

\section{Summary and Conclusions} \label{sec: summary}

Below we list the key conclusions of our analysis:

\begin{enumerate}
    \item For most of our sample, we find that the retrieved thermal structures are in agreement with with predictions from self-consistent grid based approaches (Figure \ref{fig:50_tp-profiles}). However, 18 objects have retrieved thermal profiles that are much hotter, at roughly 2$\sigma$, than the grid models. Though possibly physical in origin, it is more likely owed to the different physical and chemical assumptions in each model. Longer wavelength information, namely beyond 2.5 microns where most of the flux is, would significantly help in testing if this is the true cause of the discrepancy.
    \item In addition to the thermal profiles, the highlight of this work is the retrieval of abundance constraints for the major atmospheric molecular species [H$_2$O, CH$_4$, NH$_3$, K] with a precision of roughly 0.25dex at 1$\sigma$ (Figure \ref{fig:mol_abundance}, Table \ref{tab:retrieved_and_derived}). For a majority of our sample, we find that H$_2$O, CH$_4$, and NH$_3$ are consistent with expectations from equilibrium chemistry. There are however, a small subset of objects which do not fit these trends. There are four objects with elevated CH$_4$ and roughly ten with elevated NH$_3$ constraints. Though it may be a simple result of small number sampling, there is literature showing that objects near this temperature range may begin to show signs of elevated CH$_4$ and NH$_3$ due to non-equilibrium vertical mixing. Developing a larger sample of earlier T dwarfs (\teff $\textgreater$ 900K) requires accurately taking into consideration a realistic cloud model into our retrieval framework, but would show if this trend continues into higher temperatures. Finally we note that our sample helps confirm previous findings that the depletion of Na and K near 500-600K owed to the chemical rainout of species like enstatite and forsterite is present.
    \item Using the retrieved molecular abundances we are able to infer the atmospheric metallicity and C/O ratio. We find that the distribution of our 50 targets is broadly consistent with measurements of the local stellar FGK population. Though there is substantial scatter, there also is a slight positive correlation between atmospheric metallicity and effective temperature.
    \item We compare our constraints on both \teff and log(g) to expectations from evolutionary models and previous results of Y dwarfs (Figure \ref{fig:logg1}). For a majority of objects, we find them to be consistent with expectations of typical field age brown dwarfs. However we also find several outliers that have unphysically old and young ages. We preformed several tests in which we altered the priors on our retrieval model to ''guide" these objects to more physically realistic ages. Despite these efforts, we found that the resulting posteriors were highly non-Gaussian and were clustered near the limit of the new prior ranges. This indicates that there is some as yet undetermined feature in the spectra which favors these anomalous ages. Given that these age estimate are likely incorrect, this result highlights the importance of complimenting such retrieval studies with more traditional grid models.
   
\end{enumerate}

This uniform atmospheric retrieval analysis of 50 T dwarfs provides a large and unique dataset of retrieved thermal profiles, gravities, and atmospheric abundances which offer insight into the physical and chemical mechanisms at work in their atmospheres. This dataset now can serve as a baseline for comparisons between other stars and planets to place brown dwarfs in a holistic astronomical context.\\

\section*{Acknowledgements}
The retrieval analysis work is supported by the National Science Foundation under Grant No. 1615220. JZ acknowledges funding from a NASA FINESST grant. This research has benefited from the SpeX Prism Spectral Libraries, maintained by Adam Burgasser at this \href{http://pono.ucsd.edu/~adam/browndwarfs/spexprism}{link}. The research shown herein acknowledges use of the Hypatia Catalog Database, an online compilation of stellar abundance data as described in \cite{2014AJ....148...54H}.


\appendix

\section{Retrieved Atmospheric Properties}

\begin{figure}[b]
    \centering
    \includegraphics[width=\columnwidth]{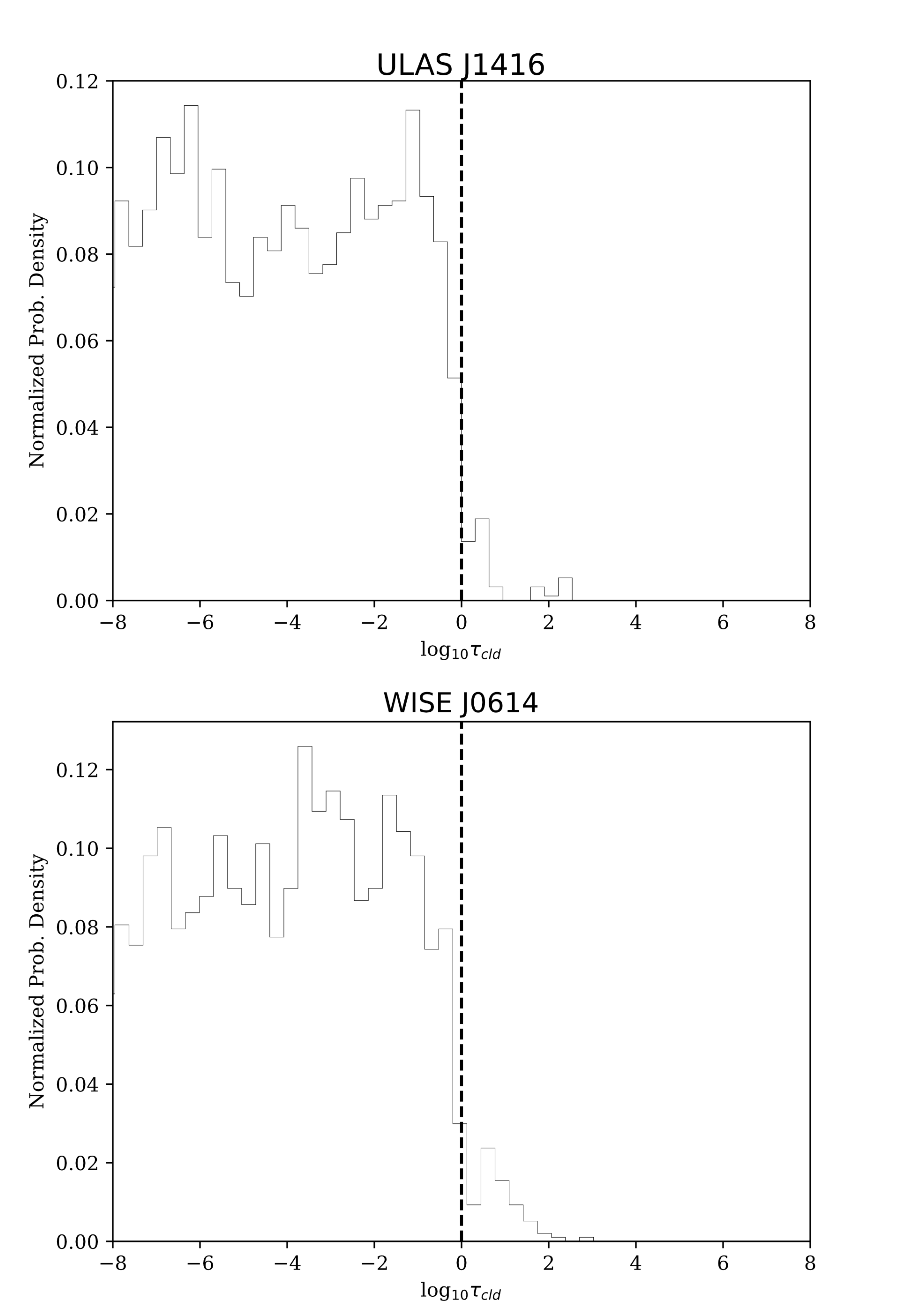}
    \caption{Cloud column optical depth ($\tau_{cld}$) posteriors for two different representative targets of our 50 target sample. The optical depth is calculated by integrating the retrieved opacity profile. The central dashed line indicates where $\tau_{cld}$=1. The histograms sharply decline as we approach this value, suggesting that the data are not affected by optically thick clouds. Profiles for all 50 objects are available in the Zenodo link in Section \ref{sec: results}.}
    \label{fig:tau_cloud}
\end{figure}

\clearpage
\startlongtable
\begin{deluxetable*}{cccccccccc}
    \tablecaption{Retrieved Atmospheric Abundances and Derived Parameters}
    \label{tab:retrieved_and_derived}
    \tablehead{Object & \teff [K]  & log(g) [cgs] & R [R$_J$] & H$_2$O [VMR] & CH$_4$ [VMR] & NH$_3$ [VMR] & K [VMR] & [M/H] & C/O}
    \startdata
        HD3651B & 715.57$^{+24.68}_{-9.65}$ & 5.07$^{+0.09}_{-0.14}$ & 1.17$^{+0.10}_{-0.10}$ & -3.33$^{+0.08}_{-0.07}$ & -3.21$^{+0.07}_{-0.07}$ & -4.54$^{+0.07}_{-0.09}$ & -7.02$^{+0.07}_{-0.09}$ & -0.04$^{+0.08}_{-0.07}$ & 0.98$^{+0.11}_{-0.11}$\\
        \midrule
        WISE-J0040 & 885.44$^{+18.84}_{-13.25}$ & 4.66$^{+0.34}_{-0.34}$ & 0.97$^{+0.10}_{-0.09}$ & -3.51$^{+0.34}_{-0.11}$ & -3.54$^{+0.15}_{-0.15}$ & -4.82$^{+0.21}_{-0.19}$ & -6.91$^{+0.09}_{-0.08}$ & -0.35$^{+0.14}_{-0.13}$ & 0.70$^{+0.11}_{-0.11}$\\
        \midrule
        WISE-J0049 & 727.28$^{+19.72}_{-9.70}$ & 5.18$^{+0.20}_{-0.33}$ & 0.81$^{+0.07}_{-0.07}$ & -3.09$^{+0.09}_{-0.12}$ & -2.96$^{+0.15}_{-0.10}$ & -4.40$^{+0.12}_{-0.19}$ & -6.99$^{+0.09}_{-0.10}$ & 0.14$^{+0.10}_{-0.14}$ & 1.01$^{+0.12}_{-0.12}$\\
        \midrule
        2MASS-J0050 & 945.64$^{+5.83}_{-12.09}$ & 5.26$^{+0.22}_{-0.19}$ & 0.75$^{+0.05}_{-0.04}$ & -3.19$^{+0.10}_{-0.12}$ & -3.07$^{+0.11}_{-0.14}$ & -4.49$^{+0.12}_{-0.13}$ & -6.37$^{+0.08}_{-0.09}$ & 0.10$^{+0.11}_{-0.12}$ & 0.98$^{+0.11}_{-0.10}$\\
        \midrule
        WISE-PAJ0123 & 872.83$^{+3.15}_{-3.97}$ & 5.15$^{+0.11}_{-0.15}$ & 1.02$^{+0.07}_{-0.06}$ & -3.51$^{+0.07}_{-0.08}$ & -3.48$^{+0.07}_{-0.08}$ & -4.59$^{+0.06}_{-0.08}$ & -6.78$^{+0.06}_{-0.06}$ & -0.31$^{+0.09}_{-0.08}$ & 0.80$^{+0.07}_{-0.06}$\\
        \midrule
        WISE-J0223 & 787.96$^{+4.11}_{-4.60}$ & 5.23$^{+0.18}_{-0.33}$ & 0.81$^{+0.09}_{-0.10}$ & -3.12$^{+0.11}_{-0.13}$ & -3.25$^{+0.09}_{-0.15}$ & -4.72$^{+0.15}_{-0.26}$ & -6.80$^{+0.09}_{-0.10}$ & -0.02$^{+0.10}_{-0.14}$ & 0.54$^{+0.08}_{-0.07}$\\
        \midrule
        WISE-J0241 & 835.78$^{+3.82}_{-4.32}$ & 4.90$^{+0.19}_{-0.22}$ & 1.06$^{+0.05}_{-0.05}$ & -3.55$^{+0.08}_{-0.08}$ & -3.51$^{+0.09}_{-0.10}$ & -4.74$^{+0.11}_{-0.12}$ & -6.93$^{+0.05}_{-0.04}$ & -0.36$^{+0.09}_{-0.09}$ & 0.82$^{+0.07}_{-0.07}$\\
        \midrule
        PSO-J043 & 699.53$^{+5.75}_{-6.22}$ & 5.22$^{+0.14}_{-0.21}$ & 0.84$^{+0.06}_{-0.06}$ & -3.22$^{+0.07}_{-0.09}$ & -3.09$^{+0.07}_{-0.10}$ & -4.47$^{+0.08}_{-0.14}$ & -6.81$^{+0.06}_{-0.07}$ & 0.02$^{+0.07}_{-0.10}$ & 1.00$^{+0.10}_{-0.11}$\\
        \midrule 
        WISE-J0325 & 897.80$^{+7.93}_{-3.47}$ & 3.54$^{+0.13}_{-0.08}$ & 0.94$^{+0.06}_{-0.05}$ & -3.77$^{+0.07}_{-0.07}$ & -4.00$^{+0.07}_{-0.06}$ & -5.47$^{+0.11}_{-0.12}$ & -6.84$^{+0.08}_{-0.08}$ & -0.69$^{+0.14}_{-0.07}$ & 0.47$^{+0.09}_{-0.05}$\\
        \midrule
        2MASS-J0415 & 675.24$^{+8.95}_{-5.38}$ & 5.10$^{+0.15}_{-0.25}$ & 0.94$^{+0.06}_{-0.06}$ & -3.24$^{+0.07}_{-0.09}$ & -3.13$^{+0.08}_{-0.12}$ & -4.53$^{+0.09}_{-0.16}$ & -6.88$^{+0.06}_{-0.06}$ & -0.02$^{+0.07}_{-0.10}$ & 0.96$^{+0.09}_{-0.10}$\\
        \midrule
        WISE-PCJ0458 & 616.50$^{+4.15}_{-9.05}$ & 4.87$^{+0.25}_{-0.45}$ & 1.15$^{+0.14}_{-0.13}$ & -3.28$^{+0.12}_{-0.19}$ & -3.40$^{+0.13}_{-0.21}$ & -4.77$^{+0.16}_{-0.23}$ & -7.32$^{+0.14}_{-0.18}$ & -0.05$^{+0.18}_{-0.21}$ & 0.53$^{+0.10}_{-0.08}$ \\
        \midrule
        UGPS-J0521 & 658.43$^{+1.21}_{-0.58}$ & 5.28$^{+0.19}_{-0.27}$ & 0.74$^{+0.04}_{-0.05}$ & -3.05$^{+0.08}_{-0.08}$ & -2.97$^{+0.09}_{-0.12}$ & -4.49$^{+0.10}_{-0.17}$ & -6.79$^{+0.07}_{-0.09}$ & 0.24$^{+0.09}_{-0.10}$ & 0.83$^{+0.09}_{-0.08}$\\
        \midrule
        WISE-J0521 & 880.60$^{+4.15}_{-3.15}$ & 4.18$^{+0.39}_{-0.28}$ & 0.85$^{+0.03}_{-0.05}$ & -3.47$^{+0.15}_{-0.08}$ & -3.54$^{+0.19}_{-0.12}$ & -5.07$^{+0.14}_{-0.21}$ & -6.79$^{+0.07}_{-0.06}$ & -0.27$^{+0.18}_{-0.10}$ & 0.63$^{+0.09}_{-0.08}$\\
        \midrule
        WISE-J0614 & 897.89$^{+4.22}_{-14.69}$ & 4.93$^{+0.12}_{-0.20}$ & 1.22$^{+0.09}_{-0.09}$ & -3.48$^{+0.07}_{-0.07}$ & -3.45$^{+0.07}_{-0.09}$ & -4.68$^{+0.08}_{-0.11}$ & -6.86$^{+0.06}_{-0.05}$ & -0.29$^{+0.07}_{-0.08}$ & 0.79$^{+0.07}_{-0.07}$\\
        \midrule
        2MASS-J0727 & 953.07$^{+3.34}_{-1.80}$ & 5.23$^{+0.17}_{-0.20}$ & 0.76$^{+0.05}_{-0.04}$ & -3.09$^{+0.10}_{-0.09}$ & -3.15$^{+0.10}_{-0.10}$ & -4.76$^{+0.10}_{-0.26}$ & -6.51$^{+0.08}_{-0.10}$ & 0.05$^{+0.10}_{-0.10}$ & 0.65$^{+0.07}_{-0.06}$\\
        \midrule
        2MASS-J0729 & 719.36$^{+6.39}_{-5.10}$ & 5.28$^{+0.15}_{-0.23}$ & 0.82$^{+0.10}_{-0.09}$ & -3.24$^{+0.09}_{-0.10}$ & -3.37$^{+0.09}_{-0.11}$ & -4.55$^{+0.10}_{-0.15}$ & -6.89$^{+0.09}_{-0.09}$ & 0.01$^{+0.15}_{-0.13}$ & 0.52$^{+0.05}_{-0.07}$\\
        \midrule
        2MASS-J0939 & 631.72$^{+5.13}_{-6.67}$ & 4.85$^{+0.29}_{-0.33}$ & 1.01$^{+0.06}_{-0.06}$ & -3.44$^{+0.12}_{-0.13}$ & -3.49$^{+0.14}_{-0.16}$ & -4.68$^{+0.15}_{-0.17}$ & -7.32$^{+0.13}_{-0.11}$ & -0.29$^{+0.13}_{-0.15}$ & 0.67$^{+0.07}_{-0.07}$\\
        \midrule
        ULAS-J1029 & 690.39$^{+6.46}_{-3.34}$ & 4.99$^{+0.19}_{-0.27}$ & 1.04$^{+0.09}_{-0.09}$ & -3.33$^{+0.09}_{-0.11}$ & -3.08$^{+0.10}_{-0.13}$ & -4.44$^{+0.10}_{-0.15}$ & -7.27$^{+0.12}_{-0.16}$ & -0.01$^{+0.11}_{-0.12}$ & 1.28$^{+0.18}_{-0.19}$\\
        \midrule
        WISE-J1039 & 745.93$^{+24.79}_{-9.72}$ & 4.58$^{+0.29}_{-0.48}$ & 1.46$^{+0.26}_{-0.23}$ & -3.62$^{+0.18}_{-0.20}$ & -3.45$^{+0.17}_{-0.23}$ & -4.79$^{+0.17}_{-0.24}$ & -7.01$^{+0.18}_{-0.17}$ & -0.33$^{+0.19}_{-0.22}$ & 1.03$^{+0.28}_{-0.21}$\\
        \midrule
        WISE-J105257 & 768.40$^{+13.94}_{-9.22}$ & 4.36$^{+0.40}_{-0.38}$ & 1.04$^{+0.07}_{-0.07}$ & -3.61$^{+0.13}_{-0.12}$ & -3.59$^{+0.18}_{-0.17}$ & -4.89$^{+0.21}_{-0.20}$ & -7.09$^{+0.06}_{-0.06}$ & -0.43$^{+0.16}_{-0.15}$ & 0.77$^{+0.13}_{-0.12}$\\
        \midrule
        2MASS-J1114 & 617.46$^{+5.43}_{-2.78}$ & 4.97$^{+0.22}_{-0.40}$ & 1.01$^{+0.11}_{-0.11}$ & -3.34$^{+0.11}_{-0.15}$ & -3.51$^{+0.11}_{-0.19}$ & -4.70$^{+0.14}_{-0.24}$ & -7.16$^{+0.11}_{-0.15}$ & -0.24$^{+0.12}_{-0.17}$ & 0.50$^{+0.08}_{-0.08}$\\
        \midrule
        2MASS-J1217 & 775.64$^{+27.09}_{-12.75}$ & 4.82$^{+0.17}_{-0.28}$ & 1.30$^{+0.12}_{-0.12}$ & -3.42$^{+0.09}_{-0.10}$ & -3.26$^{+0.09}_{-0.12}$ & -4.59$^{+0.10}_{-0.15}$ & -6.83$^{+0.09}_{-0.07}$ & -0.16$^{+0.10}_{-0.10}$ & 1.03$^{+0.15}_{-0.15}$\\
        \midrule
        WISE-J1124 & 848.44$^{+26.25}_{-9.67}$ & 4.78$^{+0.23}_{-0.26}$ & 0.99$^{+0.10}_{-0.07}$ & -3.41$^{+0.10}_{-0.11}$ & -3.45$^{+0.11}_{-0.12}$ & -4.68$^{+0.15}_{-0.15}$ & -6.75$^{+0.09}_{-0.10}$ & -0.25$^{+0.12}_{-0.11}$ & 0.83$^{+0.07}_{-0.08}$\\
        \midrule
        WISE-J1254 & 875.62$^{+6.77}_{-23.90}$ & 3.81$^{+0.46}_{-0.28}$ & 0.95$^{+0.26}_{-0.13}$ & -3.80$^{+0.19}_{-0.19}$ & -4.02$^{+0.21}_{-0.16}$ & -5.36$^{+0.28}_{-0.46}$ & -6.66$^{+0.17}_{-0.18}$ & -0.75$^{+0.19}_{-0.18}$ & 0.41$^{+0.19}_{-0.11}$\\
        \midrule
        WISE-J1257 & 758.90$^{+6.54}_{-4.72}$ & 3.77$^{+0.51}_{-0.21}$ & 0.89$^{+0.18}_{-0.09}$ & -3.75$^{+0.14}_{-0.14}$ & -3.96$^{+0.23}_{-0.13}$ & -5.75$^{+0.46}_{-1.32}$ & -6.62$^{+0.15}_{-0.17}$ & -0.67$^{+0.19}_{-0.13}$ & 0.48$^{+0.19}_{-0.09}$\\
        \midrule
        ROSS458C & 762.64$^{+6.85}_{-1.81}$ & 3.74$^{+0.33}_{-0.17}$ & 0.87$^{+0.09}_{-0.05}$ & -3.27$^{+0.11}_{-0.10}$ & -3.46$^{+0.16}_{-0.09}$ & -5.31$^{+0.22}_{-0.21}$ & -6.56$^{+0.13}_{-0.22}$ & -0.20$^{+0.14}_{-0.09}$ & 0.51$^{+0.12}_{-0.07}$\\
        \midrule
        WISE-J1322 & 775.78$^{+6.91}_{-19.53}$ & 3.87$^{+0.77}_{-0.48}$ & 1.17$^{+0.33}_{-0.22}$ & -3.73$^{+0.27}_{-0.23}$ & -3.83$^{+0.36}_{-0.25}$ & -5.38$^{+0.44}_{-0.52}$ & -6.94$^{+0.17}_{-0.17}$ & -0.59$^{+0.32}_{-0.23}$ & 0.62$^{+0.27}_{-0.18}$\\
        \midrule
        ULAS-J1416 & 616.23$^{+1.90}_{-4.07}$ & 5.12$^{+0.19}_{-0.46}$ & 0.93$^{+0.07}_{-0.06}$ & -3.47$^{+0.09}_{-0.19}$ & -3.69$^{+0.09}_{-0.21}$ & -4.76$^{+0.11}_{-0.23}$ & -7.23$^{+0.10}_{-0.21}$ & -0.38$^{+0.09}_{-0.19}$ & 0.45$^{+0.07}_{-0.06}$\\
        \midrule
        WISE-J1457 & 768.75$^{+5.59}_{-10.00}$ & 4.67$^{+0.23}_{-0.39}$ & 1.38$^{+0.22}_{-0.19}$ & -3.65$^{+0.12}_{-0.14}$ & -3.59$^{+0.12}_{-0.18}$ & -4.80$^{+0.13}_{-0.24}$ & -6.99$^{+0.10}_{-0.09}$ & -0.44$^{+0.12}_{-0.16}$ & 0.83$^{+0.15}_{-0.15}$\\
        \midrule
        Gliese 570D & 789.90 $^{+4.05}_{-12.67}$ & 5.14$^{+0.14}_{-0.26}$ & 0.95$^{+0.09}_{-0.11}$ & -3.27$^{+0.09}_{-0.09}$ & -3.25$^{+0.08}_{-0.12}$ & -4.58$^{+0.10}_{-0.21}$ & -6.81$^{+0.08}_{-0.07}$ & -0.02$^{+0.10}_{-0.10}$ & 0.77$^{+0.08}_{-0.10}$\\
        \midrule
        PSO-J224 & 892.88$^{+8.55}_{-25.81}$ & 3.64$^{+0.25}_{-0.14}$ & 0.91$^{+0.07}_{-0.06}$ & -3.86$^{+0.09}_{-0.09}$ & -3.92$^{+0.12}_{-0.08}$ & -5.25$^{+0.13}_{-0.11}$ & -6.89$^{+0.08}_{-0.08}$ & -0.62$^{+0.16}_{-0.10}$ & 0.68$^{+0.09}_{-0.08}$\\
        \midrule
        SDSS-1504 & 872.47$^{+5.04}_{-9.67}$ & 4.98$^{+0.12}_{-0.20}$ & 1.22$^{+0.08}_{-0.07}$ & -3.55$^{+0.06}_{-0.08}$ & -3.45$^{+0.06}_{-0.10}$ & -4.63$^{+0.07}_{-0.11}$ & -6.93$^{+0.05}_{-0.05}$ & -0.32$^{+0.07}_{-0.09}$ & 0.91$^{+0.08}_{-0.08}$\\
        \midrule
        2MASS-J1553 & 900.88$^{+2.03}_{-1.83}$ & 4.56$^{+0.20}_{-0.20}$ & 1.13$^{+0.06}_{-0.06}$ & -3.38$^{+0.08}_{-0.08}$ & -3.49$^{+0.09}_{-0.09}$ & -5.06$^{+0.13}_{-0.21}$ & -6.60$^{+0.06}_{-0.14}$ & -0.27$^{+0.09}_{-0.09}$ & 0.58$^{+0.05}_{-0.06}$\\
        \midrule
        SDSS-J16283 & 916.45$^{+6.58}_{-10.41}$ & 4.91$^{+0.31}_{-0.40}$ & 0.83$^{+0.07}_{-0.06}$ & -3.29$^{+0.12}_{-0.13}$ & -3.28$^{+0.15}_{-0.18}$ & -4.76$^{+0.20}_{-0.22}$ & -6.58$^{+0.09}_{-0.09}$ & -0.11$^{+0.14}_{-0.16}$ & 0.73$^{+0.13}_{-0.10}$\\
        \midrule
        WISE-J1653 & 687.22$^{+15.06}_{-12.25}$ & 4.87$^{+0.28}_{-0.44}$ & 1.03$^{+0.12}_{-0.13}$ & -3.27$^{+0.13}_{-0.16}$ & -3.30$^{+0.13}_{-0.21}$ & -4.77$^{+0.19}_{-0.29}$ & -7.01$^{+0.11}_{-0.10}$ & -0.11$^{+0.14}_{-0.19}$ & 0.66$^{+0.12}_{-0.11}$\\
        \midrule
        WISE-PAJ1711 & 805.55$^{+2.76}_{-1.80}$ & 4.80$^{+0.27}_{-0.43}$ & 1.17$^{+0.23}_{-0.15}$ & -3.45$^{+0.14}_{-0.18}$ & -3.40$^{+0.14}_{-0.21}$ & -4.74$^{+0.14}_{-0.24}$ & -7.05$^{+0.11}_{-0.12}$ & -0.24$^{+0.14}_{-0.19}$ & 0.82$^{+0.15}_{-0.14}$\\
        \midrule
        WISE-PAJ1741 & 646.58$^{+4.26}_{-4.58}$ & 4.64$^{+0.18}_{-0.18}$ & 0.84$^{+0.03}_{-0.04}$ & -3.29$^{+0.08}_{-0.06}$ & -3.19$^{+0.09}_{-0.08}$ & -4.62$^{+0.09}_{-0.10}$ & -7.33$^{+0.09}_{-0.09}$ & -0.08$^{+0.08}_{-0.07}$ & 0.95$^{+0.07}_{-0.08}$\\
        \midrule
        WISE-J1813 & 706.42$^{+4.04}_{-4.45}$ & 4.51$^{+0.13}_{-0.20}$ & 2.04$^{+0.19}_{-0.35}$ & -3.45$^{+0.11}_{-0.08}$ & -3.29$^{+0.07}_{-0.10}$ & -4.76$^{+0.09}_{-0.27}$ & -7.23$^{+0.13}_{-0.11}$ & -0.19$^{+0.09}_{-0.08}$ & 1.02$^{+0.14}_{-0.23}$\\
        \midrule
        WISEPA J1852 & 915.71$^{+14.64}_{-5.53}$ & 5.12$^{+0.22}_{-0.28}$ & 0.80$^{+0.10}_{-0.05}$ & -3.16$^{+0.12}_{-0.15}$ & -3.06$^{+0.13}_{-0.15}$ & -4.62$^{+0.14}_{-0.20}$ & -6.38$^{+0.09}_{-0.13}$ & 0.06$^{+0.13}_{-0.16}$ & 0.93$^{+0.16}_{-0.12}$\\
        \midrule
        WISE-PAJ1959 & 754.21$^{+17.80}_{-6.23}$ & 3.74$^{+0.32}_{-0.18}$ & 0.91$^{+0.07}_{-0.06}$ & -3.64$^{+0.08}_{-0.07}$ & -3.76$^{+0.07}_{-0.13}$ & -5.16$^{+0.15}_{-0.10}$ & -7.00$^{+0.09}_{-0.09}$ & -0.53$^{+0.11}_{-0.08}$ & 0.58$^{+0.11}_{-0.07}$\\
        \midrule
        WISE-J2000 & 753.40$^{+3.29}_{-5.34}$ & 4.76$^{+0.39}_{-0.27}$ & 0.90$^{+0.06}_{-0.05}$ & -3.31$^{+0.12}_{-0.10}$ & -3.34$^{+0.17}_{-0.13}$ & -4.80$^{+0.21}_{-0.17}$ & -7.05$^{+0.07}_{-0.06}$ & -0.17$^{+0.15}_{-0.11}$ & 0.70$^{+0.09}_{-0.08}$\\
        \midrule
        WISE-PCJ2157 & 792.95$^{+12.10}_{-9.11}$ & 5.22$^{+0.15}_{-0.25}$ & 0.83$^{+0.05}_{-0.05}$ & -3.27$^{+0.07}_{-0.08}$ & -3.22$^{+0.07}_{-0.12}$ & -4.57$^{+0.10}_{-0.15}$ & -6.89$^{+0.06}_{-0.06}$ & -0.02$^{+0.07}_{-0.10}$ & 0.85$^{+0.06}_{-0.07}$\\
        \midrule
        WISE-PCJ2209 & 809.34$^{+5.18}_{-4.03}$ & 5.00$^{+0.21}_{-0.30}$ & 0.90$^{+0.05}_{-0.04}$ & -3.37$^{+0.08}_{-0.12}$ & -3.54$^{+0.10}_{-0.14}$ & -4.79$^{+0.13}_{-0.19}$ & -6.95$^{+0.06}_{-0.06}$ & -0.28$^{+0.10}_{-0.13}$ & 0.50$^{+0.05}_{-0.05}$\\
        \midrule
        WISE-J2213 & 852.44$^{+11.78}_{-11.13}$ & 4.65$^{+0.33}_{-0.28}$ & 0.96$^{+0.09}_{-0.06}$ & -3.50$^{+0.10}_{-0.09}$ & -3.57$^{+0.14}_{-0.13}$ & -4.86$^{+0.20}_{-0.16}$ & -6.73$^{+0.07}_{-0.08}$ & - 0.36$^{+0.12}_{-0.11}$ &  0.63$^{+0.12}_{-0.08}$\\
        \midrule
        WISE-J2226 & 868.13$^{+18.55}_{-13.44}$ & 5.27$^{+0.18}_{-0.27}$ & 0.78$^{+0.09}_{-0.06}$ & -3.02$^{+0.11}_{-0.15}$ & -2.98$^{+0.11}_{-0.14}$ & -4.85$^{+0.24}_{-0.37}$ & -6.48$^{+0.09}_{-0.14}$ & 0.16$^{+0.11}_{-0.14}$ & 0.82$^{+0.14}_{-0.10}$\\
        \midrule
        WISE-J2255 & 720.23$^{+3.48}_{-7.03}$ & 5.22$^{+0.15}_{-0.23}$ & 0.89$^{+0.11}_{-0.10}$ & -3.09$^{+0.12}_{-0.11}$ & -2.94$^{+0.10}_{-0.11}$ & -4.51$^{+0.11}_{-0.18}$ & -6.90$^{+0.13}_{-0.12}$ & 0.16$^{+0.11}_{-0.11}$ & 1.04$^{+0.17}_{-0.15}$\\
        \midrule
        WISE-J2319 & 714.16$^{+5.82}_{-2.84}$ & 4.87$^{+0.59}_{-0.64}$ & 0.62$^{+0.09}_{-0.07}$ & -3.44$^{+0.25}_{-0.25}$ & -3.43$^{+0.28}_{-0.30}$ & -4.61$^{+0.29}_{-0.32}$ & -6.99$^{+0.16}_{-0.15}$ & -0.24$^{+0.26}_{-0.28}$ & 0.74$^{+0.15}_{-0.13}$\\
        \midrule
        ULAS-J2321 & 750.05$^{+8.29}_{-8.48}$ & 5.23$^{+0.11}_{-0.17}$ & 0.90$^{+0.06}_{-0.07}$ & -3.14$^{+0.07}_{-0.07}$ & -2.96$^{+0.07}_{-0.08}$ & -4.36$^{+0.07}_{-0.11}$ & -6.74$^{+0.07}_{-0.07}$ & 0.13$^{+0.07}_{-0.08}$ & 1.12$^{+0.11}_{-0.11}$\\
        \midrule
        WISE-J2340 & 904.21$^{+5.96}_{-5.98}$ & 4.72$^{+0.17}_{-0.23}$ & 1.47$^{+0.14}_{-0.11}$ & -3.58$^{+0.09}_{-0.10}$ & -3.56$^{+0.09}_{-0.11}$ & -4.87$^{+0.10}_{-0.15}$ & -6.73$^{+0.08}_{-0.08}$ & -0.40$^{+0.10}_{-0.11}$ & 0.78$^{+0.09}_{-0.08}$\\
        \midrule
        WISE-J2348 & 921.60$^{+1.08}_{-9.23}$ & 4.94$^{+0.29}_{-0.32}$ & 0.77$^{+0.20}_{-0.07}$ & -3.28$^{+0.17}_{-0.17}$ & -3.20$^{+0.17}_{-0.15}$ & -4.74$^{+0.19}_{-0.22}$ & -6.51$^{+0.15}_{-0.27}$ & -0.07$^{+0.17}_{-0.16}$ & 0.87$^{+0.23}_{-0.13}$\\
        \midrule
    \enddata
\end{deluxetable*}

\begin{figure}
    \centering
    \addtolength{\leftskip}{-0.5cm}
    \includegraphics[width=1.0\textwidth]{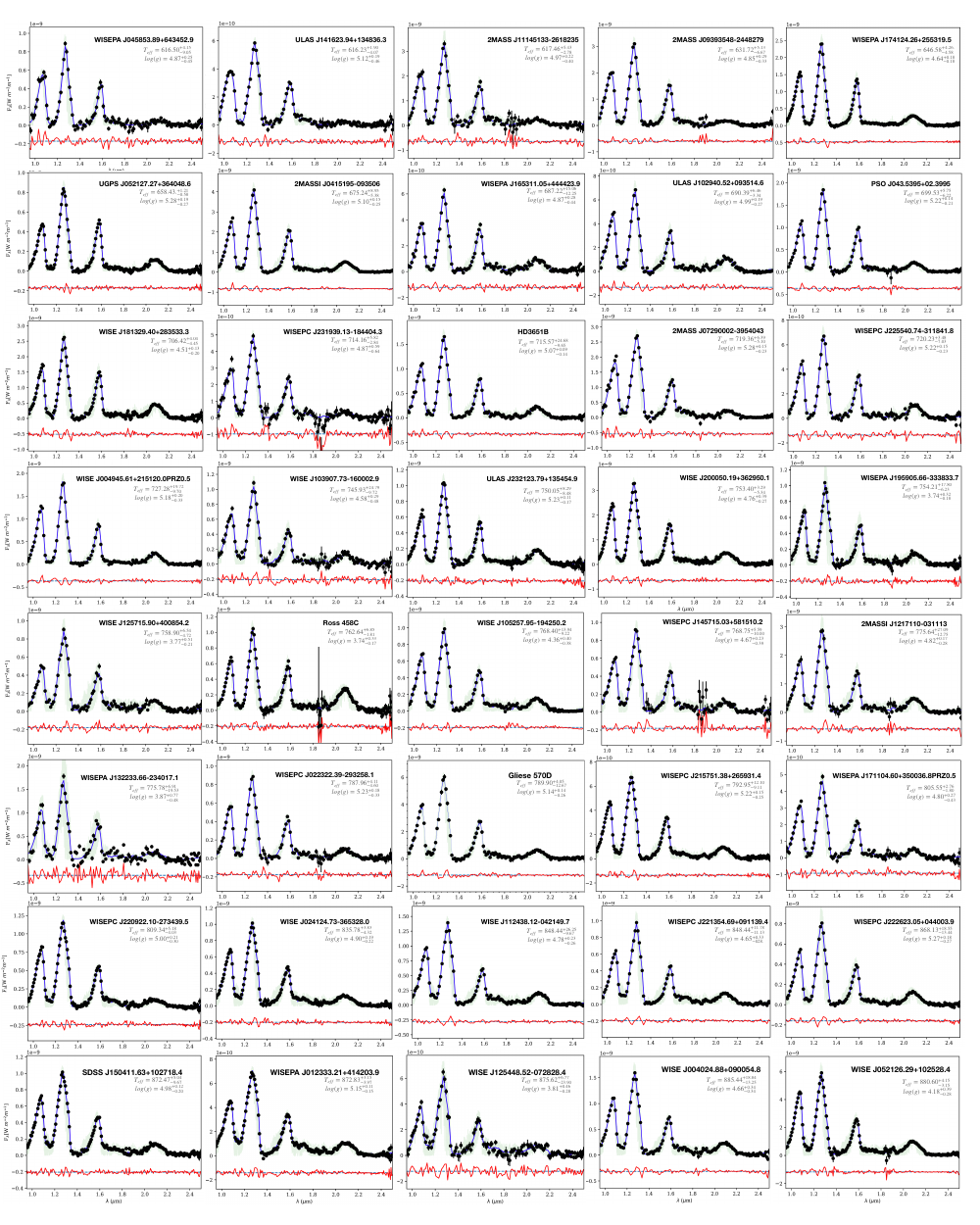}
    
    \label{fig:50_spec}
\end{figure}

\begin{figure*}
    \centering
    \addtolength{\leftskip}{-0.5cm}
    \includegraphics[width=1.0\textwidth]{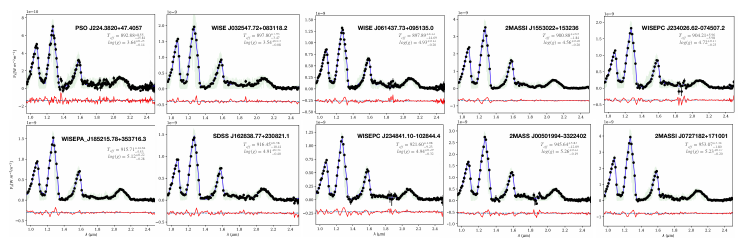}
    \caption{A continuation of Figure \ref{fig:spectra} showing the SpeX spectra (black), best-fit spectra (blue) and residuals (red) for all objects in our sample.} 
    \label{fig:50_spec}
\end{figure*}

\begin{figure}[b]
    \centering
    \addtolength{\leftskip}{-1.0cm}
    \includegraphics[width=1.0\textwidth]{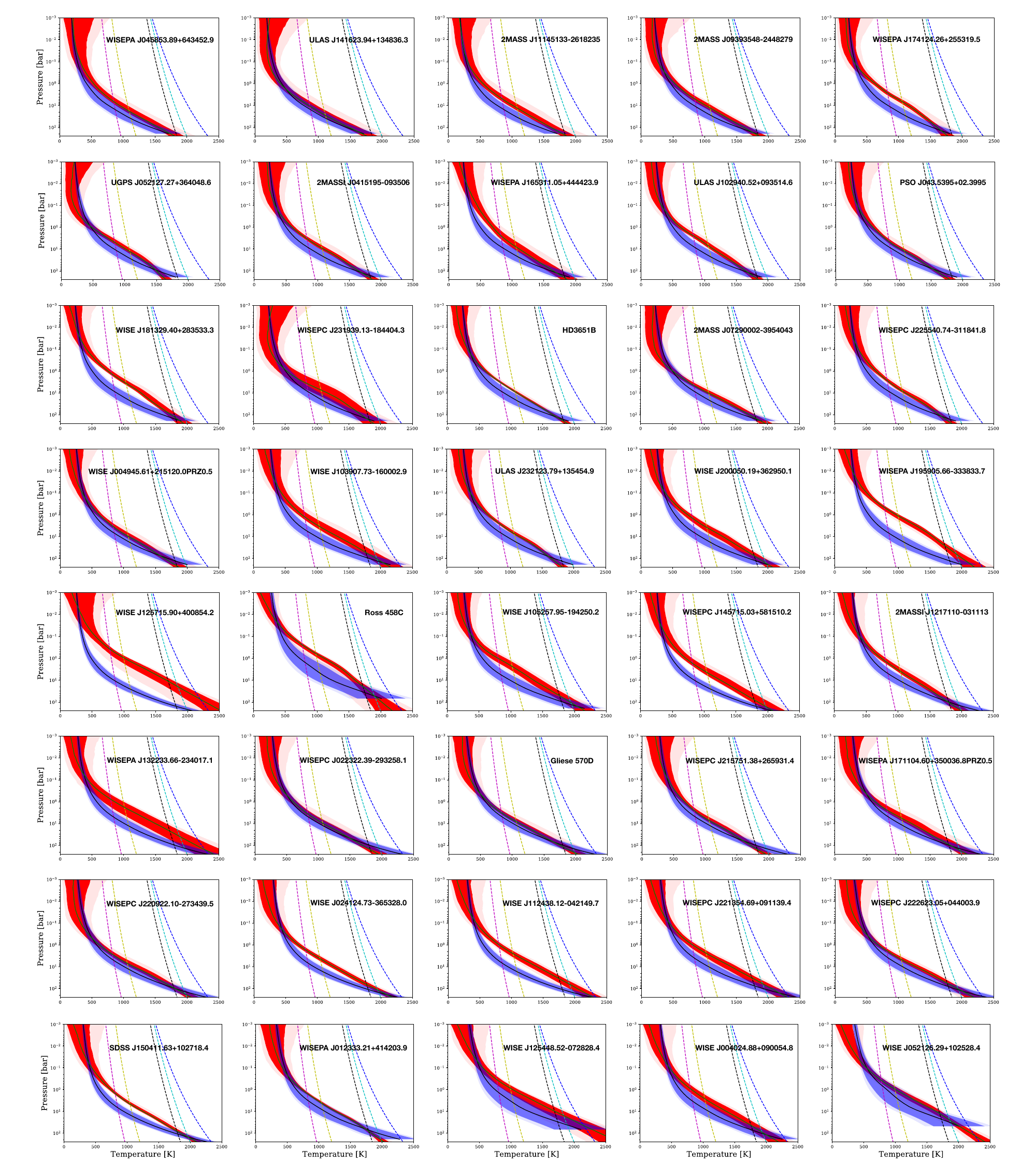}
    \label{fig:50_tp-profiles}
\end{figure}

\begin{figure*}
    \centering
    \addtolength{\leftskip}{-0.8cm}
    \includegraphics[width=1.0\textwidth]{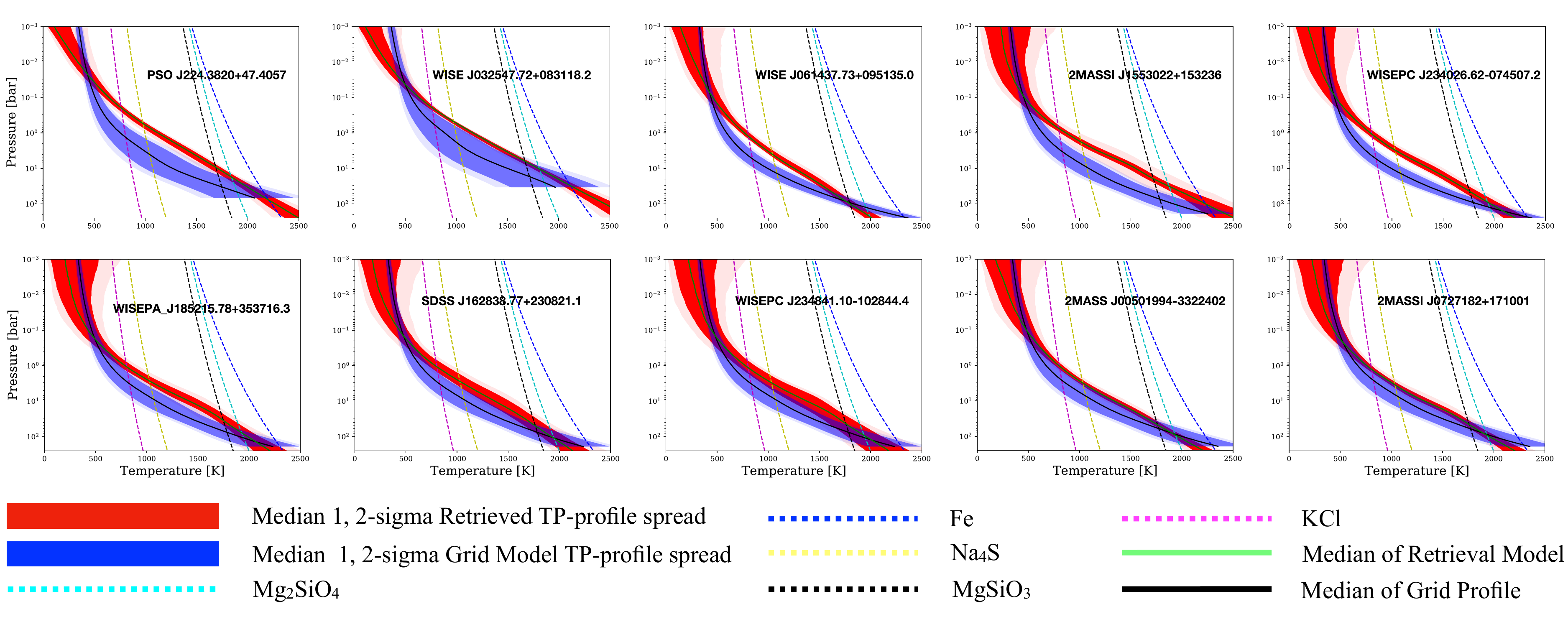}
    \caption{A continuation of Figure \ref{fig:tp_profiles} showing the retrieved (red) and grid-model (blue) PT profiles, overlaid with relevant condensate curves, for all objects in our sample.} 
    \label{fig:50_tp-profiles}
\end{figure*}

\bibliographystyle{aasjournal}
\clearpage
\bibliography{main}{}
\end{document}